\documentclass[structabstract]{aa}
\pdfoutput=1
\usepackage{txfonts}
\usepackage{pictex}
\usepackage{latexsym}
\usepackage{graphics}
\usepackage{afterpage}
\usepackage{graphicx}

\begin{document}

% \thesaurus{11(11.03.4; 11.06.2; 11.07.1; 11.16.1)}

\title
{
Ultraviolet variability of quasars: dependence\\
 on the accretion rate\thanks{The catalogue of
quasars is only available at the CDS via
anonymous ftp to cdsarc.u-strasbg.fr (130.79.128.5) or via
http://cdsarc.u-strasbg.fr/viz-bin/qcat/A+A/vol/page}
}

\author{H. Meusinger      \inst{1}
	 \and 
	 V. Weiss    \inst{1}
	 }
	 
    \institute{
	 Th\"uringer Landessternwarte Tautenburg, Sternwarte 5, D--07778
         Tautenburg, Germany,
	       e-mail: meus@tls-tautenburg.de
        }
	
\date{Received / Accepted }

\abstract {}
   {
   Although the variability in the ultraviolet and optical domain is one of the major characteristics of quasars, the dominant
   underlying mechanisms are still poorly understood. There is a broad consensus on the relationship between the strength of the
   variability and such quantities as time-lag, wavelength, luminosity, and redshift. However, evidence on a dependence on the
   fundamental parameters of the accretion process is still inconclusive. This paper is focused on the correlation between the
   ultraviolet quasar long-term variability and the accretion rate.
   }
   {
   We compiled a catalogue of about 4000 quasars including individual estimators for the variability strength derived from the
   multi-epoch photometry in the SDSS Stripe 82, virial black hole masses $M$ derived from the \ion{Mg}{ii} line, and
   mass accretion rates $\dot{M}$ from the Davis-Laor scaling relation.
   Several statistical tests were applied to evaluate the correlations of the variability with luminosity, mass,
   Eddington ratio, and accretion rate.
   }
   {
   We confirm the existence of significant anti-correlations between the variability estimator $V$ and the accretion rate
   $\dot{M}$, the Eddington ratio $\varepsilon$, and the bolometric luminosity $L_{\rm bol}$, respectively.
   The Eddington ratio is tightly correlated with $\dot{M}$. A weak, statistically not significant positive trend is
   indicated for the dependence of $V$ on $M$. 
   As a side product, we find a strong correlation of the radiative efficiency $\eta$ with $M$ in our sample.
   We show via numerical simulations that this trend is most likely produced by selection effects in combination with
   the mass errors and the use of the scaling relation for $\dot{M}$.
   The anti-correlations of $V$ with $\dot{M}$, $\varepsilon$, and $L_{\rm bol}$ cannot be explained in such a way.
   The strongest anti-correlation is found between $V$ and $\dot{M}$.
   However, it is difficult to decide which of the quantities $L, \varepsilon$, and $\dot{M}$  is intrinsically correlated
   with $V$ and which of the observed correlations of $V$ are produced by the $L - \varepsilon - \dot{M}$ relation.
   A $V-\dot{M}$ anti-correlation is
   qualitatively expected for the strongly inhomogeneous accretion disks.
   We argue that the observed amplitudes of the variability at far UV wavelengths, the stochastic nature of variability,
   and the variability time-scales are not adequately explained by the simple multi-temperature black-body model of a
   standard disk and suggest to check whether the strongly inhomogeneous disk model is capable of reproducing these
   observations better.
   }
   {}      

\titlerunning{Quasar variability and accretion rate}
\authorrunning{H. Meusinger \& V. Weiss}

\maketitle

%%%%%%%%%%%%%%%%%%%%%%%%%%%%%%%%%%%%%%%%%%%%%%%%%%%%%%%%%%%%%%%%%%%%
%
%
\section{Introduction}
%
%
%%%%%%%%%%%%%%%%%%%%%%%%%%%%%%%%%%%%%%%%%%%%%%%%%%%%%%%%%%%%%%%%%%%%

It has long been suggested that the power for the large luminosities 
($10^{11} \la L_{\rm bol}/L_\odot \la 10^{14}$) of quasars is provided by 
mass accretion onto supermassive black holes (BHs)
(Salpeter \cite{Salpeter64}; 
Lynden-Bell \cite{Lynden-Bell69};
Rees \cite{Rees84}).
In the steep gravitational potential of a BH with the mass $M$,
the gravitational energy of the accreted matter is transformed into
radiation with an efficiency $\eta$.
The relative growth rate of the BH can be described by the Eddington ratio
$\varepsilon \equiv L_{\rm bol}/L_{\rm Edd} = \dot{M}/\dot{M}_{\rm Edd}$,
where $L_{\rm Edd}$ and $\dot{M}_{\rm Edd}$ are the luminosity and the accretion rate,
respectively, for the critical stable case where the inward gravitational pressure of the
accretion flow is just balanced out by the outward pressure of the radiation
flow. According to the standard picture
(Shakura \& Sunyaev \cite{Shakura73}; 
Novikov \& Thorne \cite{Novikov73};
Shields \cite{Shields78};
Frank et al. \cite{Frank02};
Alexander \& Hickox \cite{Alexander12}),
the outward angular momentum transfer of the 
accreting matter leads to the formation of an optically thick and
geometrically thin accretion disk (AD), provided that the accretion rate 
is not too small.
The particles in the AD are caused to lose angular momentum because of friction of
adjacent layers. In the simplest models, the released gravitational energy is emitted locally as
black-body radiation at the local effective temperature and the UV/optical
spectrum is thus thought to be comprised of multi-temperature
black-body components.

While the standard model basically agrees with many observed properties of X-ray
binaries and AGNs, several observations of the spectral energy distribution and the
variability of quasars imply modifications beyond the simplest models
(e.g.,
Koratkar \& Blaes \cite{Koratkar99}; 
Agol \& Krolik \cite{Agol00}; 
Lawrence \cite{Lawrence05};
Bonning et al. \cite{Bonning07};
Davis et al. \cite{Davis07};
MacLeod et al. \cite{MacLeod10};
Schmidt et al. \cite{Schmidt12}).
In the last years, microlensing observations of quasars confirmed the model prediction
that the AD size increases with $M$. However, the disks appear to be about 5 times
larger than predicted
(Pooley et al. \cite{Pooley07};
Morgan et al. \cite{Morgan10};
Blackburne et al. \cite{Blackburne11};
Jim\'enez-Vicente et al. \cite{Jimenez12}).
Dexter \& Agol (\cite{Dexter11}; see also Dexter \& Quataert \cite{Dexter12})
argued that a modification of the standard model by the additional assumption of
local inhomogeneities in the AD can solve the size problem while matching the
stochastic properties of quasar variability.

Variability is an important diagnostic of quasar geometry and
statistical relations between empirical variability estimators and other properties
were frequently invoked to constrain the physical processes in quasars
(e.g., Kawaguchi et al. \cite{Kawaguchi98};
Tr\`evese et al. \cite{Trevese01};
Kelly et al. \cite{Kelly09}).
Numerous studies have investigated the relations between the variability amplitudes
on the one hand and time-lag, rest-frame wavelength, luminosity, and redshift, 
respectively, on the other hand. 
Most of the earlier studies
(Angione \cite{Angione73};
Uomoto et al. \cite{Uomoto76};
Bonoli et al. \cite{Bonoli79};
Netzer \& Scheffer \cite{Netzer83};
Pica \& Smith \cite{Pica83};
Cutri et al. \cite{Cutri85};
Tr\`evese et al. \cite{Trevese89};
Cristiani et al. \cite{Cristiani90};
Giallongo et al. \cite{Giallongo91};
Kinney et al. \cite{Kinney91};
Hook et al. \cite{Hook94};
Meusinger et al. \cite{Meusinger94};
V\'eron \& Hawkins \cite{Veron95};
Giveon \cite{Giveon99};
Hawkins \cite{Hawkins00};
Helfand \cite{Helfand01})
confirmed a correlation of the variability amplitudes with
the time-lag and anti-correlations with luminosity, redshift, and
rest-frame wavelength. However, with the relatively small quasar 
samples and the limited number of observation epochs
it was difficult to disentangle the effects
of the different parameters. Substantial improvement has been achieved with the
advance of large surveys. 
Over approximately the last decade, large and well-defined
quasar samples have provided a solid base for the statistical investigation of quasar variability
(Vanden Berk et al. \cite{VandenBerk04};
de Vries et al. \cite{deVries05};
Rengstorf et al. \cite{Rengstorf06};
Wold et al. \cite{Wold07};
Wilhite et al. \cite{Wilhite05}, \cite{Wilhite08};
Bauer et al. \cite{Bauer09};
Ai et al. \cite{Ai10};
MacLeod et al. \cite{MacLeod10};
Schmidt et al. \cite{Schmidt10}, \cite{Schmidt12};
Zuo et al. \cite{Zuo12}).

In our previous study (Meusinger et al. \cite{Meusinger11}; hereafter Paper\,I),
we exploited the Light-Motion Curve Catalogue (LMCC; Bramich et al. \cite{Bramich08})
of 3.7 million objects with multi-epoch photometry from the Stripe 82 (S82) of the
Sloan Digital Sky Survey Data Release 7 (SDSS DR7; Abazajian et al. \cite{Abazajian09})
to analyse the light curves for about 9000 quasars in the five SDSS bands. Because of the increase of the
variability with the time-lag, we created a variability
indicator that is related to the same length of the rest-frame time-lag interval for all quasars.
We confirmed the anti-correlations of $V$ with luminosity and redshift and could show
that the observed $V-z$ anti-correlation is caused by the combined effect of the $V-L$ and $L-z$
relations. From the analysis of the ratios of the lag-corrected variability indicators $V_j$
in adjacent photometric bands $j$ we found that the variability spectrum can be described by
$\sigma_{\rm F} \propto \lambda^{-2}$ \ ($\sigma_{\rm F}$: rms of the flux
$F_{\lambda}$), in good agreement with the result reported by Wilhite et al.
(\cite{Wilhite05}) for about $300$ quasars with SDSS spectra from several epochs.
In principle, this behaviour can be accounted for
by a standard AD that is varying from one steady state to another by changing its mean
accretion rate (Pereyra et al. \cite{Pereyra06};
Li \& Cao \cite{Li08}). However, this model cannot explain the observed variability time-scales.
Moreover, the flat power spectrum of the variations suggests a
stochastic nature of variability rather than coherent variations of the entire disk
(Kelly et al. \cite{Kelly09}; MacLeod et al. \cite{MacLeod10}).
Recently, Dexter \& Agol (\cite{Dexter11}) argued that quasar variability is probably the added
effect of many, independently varying regions of the AD.

The dependence of the variability amplitudes on $\varepsilon, M$,
and $\dot{M}$ has been the subject of several previous studies.
Wold et al. (\cite{Wold07}) estimated virial BH masses of about $100$ SDSS quasars
from the H$\beta$ line. They found a significant correlation between $M$
and the variability amplitudes from the QUEST1 Variability Survey but did not
reproduce the $V-L$ anti-correlation. Wilhite et al. (\cite{Wilhite08}) criticised that
the $V-M$ correlation found by Wold et al. would be more convincing
for a larger sample. They studied a sample of about $2500$ SDSS quasars with a
sufficiently high redshift ($z>1.69$) such that $M$ can be estimated from
the \ion{C}{iv} line profile. By comparing the ensemble variability of several
sub-samples, Wilhite et al. reproduced both the
$V-L$ anti-correlation and the $V-M$ correlation and identified the Eddington ratio
as a possible driver of quasar variability, perhaps related to the quasar's accretion
efficiency. These results were confirmed both by the investigation of the ensemble
variability of $23000$ quasars from the Palomar QUEST Survey by Bauer et al.
(\cite{Bauer09}) and the study of the individual variability indicators of
about $300$  lower-redshift ($z=0.3-0.8$) AGNs from the SDSS S82 by Ai et al. (\cite{Ai10}).
Recently, Zuo et al. (\cite{Zuo12}) studied the individual
variability estimators of more than 7000 SDSS S82 quasars and found that the
anti-correlations of $V$ with $L$ and the Eddington ratio are confirmed but that the
$V-M$ relationship is uncertain. They concluded that other physical mechanisms may
still need to be considered.

In the present paper, we search for correlations between the UV
variability and the primary quantities determining the accretion process: accretion
rate, Eddington ratio, BH mass, and luminosity for a large number of SDSS S82 quasars.
We use the individual variability indicators from Paper 1 in combination with the
Catalog of Quasar Properties from Sloan Digital Sky Survey Data Release 7
(Shen et al. \cite{Shen11}). Accretion rates are computed following Davis \& Laor (\cite{Davis11})
where it is assumed that the properties of the standard AD model are basically correct.
In Sect.\,\ref{sec:continuum}, we argue that the spectral energy distribution predicted by
the standard model is in good agreement with the observed quasar composite spectra from
the far-UV (FUV) to the near-IR (NIR).  The quasar sample of the present study and its basic properties
are presented in Sect.\,\ref{sec:sample}. In Sect.\,\ref{sec:slope-luminosity}, we discuss
the slope-luminosity relations (quasar-to-quasar and intrinsic, respectively) of
the quasars in our sample. The statistical analysis of the observed relations bet\-ween the
variability and the accretion parameters is the subject of Sect.\,\ref{sec:correlations}. Finally,
summary and conclusions are given in Sect.\,\ref{sec:conclusions}.\\

%%%%%%%%%%%%%%%%%%%%%%%%%%%%%%%%%%%%%%%%%%%%%%%%%%%%%%%%%%%%%%%%%%%%%%%%%%%%%%%%%%%%%%%%%%%%%%%%%%%%%%%%%%%%%%%%%%
%
%
\section{Quasar continuum from the extreme UV to the near-IR}\label{sec:continuum}
%
%
%%%%%%%%%%%%%%%%%%%%%%%%%%%%%%%%%%%%%%%%%%%%%%%%%%%%%%%%%%%%%%%%%%%%%%%%%%%%%%%%%%%%%%%%%%%%%%%%%%%%%%%%%%%%%%%%%%

% Fig 1a,b
\begin{figure*}[bhtp] 
\includegraphics[width=9cm]{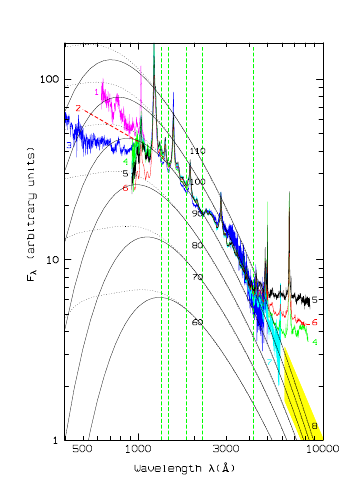}
\includegraphics[width=9cm]{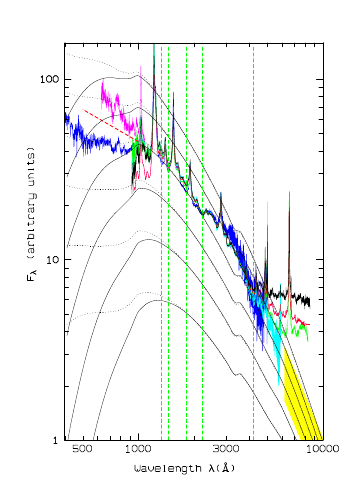}
\caption{
{\it Left:} Observed quasar composite spectra compared with the predicted continuum in the MTBB model with
different temperature parameters (labels in units of $10^3$\,K)
for a Schwarzschild black hole (solid) and for a maximum rotating Kerr BH (dotted).
{\it Right:} The same as left, but for corrected model curves of the CMTBB model. The dashed vertical lines
indicate the positions of continuum windows.
}
{\footnotesize References: 1 - Scott et al. (\cite{Scott04}); 2 - Shull et al. (\cite{Shull12}); 3 - Telfer et al. (\cite{Telfer02});
4 - Brotherton et al. (\cite{Brotherton01}); 5 - Meusinger et al. (\cite{Meusinger11}); 
6 - Vanden Berk et al. (\cite{VandenBerk04});
7 - Francis et al. (\cite{Francis91}); 8 - Kishimoto et al. (\cite{Kishimoto08})}
\label{fig:AD_spectra}
\end{figure*}
%qso_sdss_s82/variable_acc_rate/mean_spectrum/model_10_1_und_3_0p75/plot_AD_spectrum_with_composits_new_18jan13_mod_10jun13.prg
%qso_sdss_s82/variable_acc_rate/mean_spectrum/model_10_1_und_3_0p75/plot_AD_spectrum_Hubeny_corr_with_composits_new_18jan13_mod_10jun13.prg

For BH masses of $M \approx 10^8...10^9 M_\odot$, thought to be typical for quasars,
the peak effective temperature of a thermal accretion disk (AD) is expected to be
$10^4...10^5$ K. This is roughly consistent with the high continuum radiation
of quasars in the UV (Shang et al. \cite{Shang11}). This so-called big blue bump
is thus often identified with the putative AD.
The observed quasar continuum from the far UV (FUV) to the near IR (NIR) is broadly 
consistent with the predictions from the standard model
(e.g., Malkan \cite{Malkan83};
Czerny \& Elvis \cite{Czerny87};
Laor \cite{Laor90};
Pereyra et al. \cite{Pereyra06};
Kishimoto et al. \cite{Kishimoto08}).

Figure\,\ref{fig:AD_spectra} displays various composite spectra with high signal-to-noise ratio ($S/N$):
657 quasars from the  FIRST Bright Quasar Survey (FBQS; Brotherton et al. \cite{Brotherton01}),
718 quasars from the Large Bright Quasar Survey (LBQS; Francis et al. \cite{Francis91}\footnote{See
http://www.mso.anu.edu.au/pfrancis/composite/ for an updated and improved version.}),
and 2200 quasars from the Sloan Digital Sky Survey (SDSS; Vanden Berk et al. \cite{VandenBerk01}). Also plotted
is the composite from the 8744 SDSS S82 quasars in our basic sample (Paper 1).
All spectra were normalised at 2200\,\AA. The FUV to NIR continuum flux can be approximated
by a power law
\begin{equation}\label{eqn:power-law}
 F_\lambda \propto \lambda^{\alpha_\lambda}
\end{equation}
with $\alpha_\lambda = -1.54$ in the UV at $\lambda \ga 1000$\AA\ (Vanden Berk et al. \cite{VandenBerk01}).
The slope slightly steepens for increasing $\lambda$. Kishimoto et al. (\cite{Kishimoto08}) derived 
$\alpha_\lambda = -2.44$ in the NIR. This value is close to the expected ``characteristic slope''
$\alpha_\lambda = -2.33$ ($\alpha_\nu = 1/3$;
Lynden-Bell \cite{Lynden-Bell69})\footnote{I.e., the spectral slope at frequencies
$\nu \ll k_{\rm B} T_{\rm in}/h$, where $T_{\rm in}$ is the temperature at the inner edge of the AD.
For typical quasars this condition is fulfilled in the IR.}.

Redwards of H$\beta$, most composites rise above the extrapolation from the UV.
The substantial differences between different composites in the optical
are most likely the result of different contributions from the host galaxies
owing to different luminosity cuts of the samples.
For example, Vanden Berk et al. (\cite{VandenBerk01}) estimated a stellar contribution of about $7$\% at
4000\AA\ for the SDSS composite from the presence of stellar absorption lines.
In the IR, the AD is expected to be revealed in the polarised light where the polarisation
is interpreted as caused by scattering of the disk spectrum by electrons interior to the broad line region
(Kishimoto et al. \cite{Kishimoto08}).

The investigation of the extreme UV (EUV) continuum in the optical spectra of high redshift ($z \ga 2.5$) quasars
is complicated by the dense intergalactic Ly\,$\alpha$ absorption along the line of sight (Ly\,$\alpha$ forest).
Available composites derived from space-borne UV spectra of low-redshift quasars observed either with
the {\it HST Faint Object Spectrograph} (Telfer et al. \cite{Telfer02}),
the {\it HST Cosmic Origins Spectrograph} (Shull et al. \cite{Shull12}),
or with the {\it Far Ultraviolet Spectroscopic Explorer} (Scott et al. \cite{Scott04})
lead to somewhat discrepant results, probably because of the small numbers of quasars in combination with large
quasar-to-quasar variations and luminosity effects.

In the multi-temperature black-body (MTBB) model, the entire radiation flux $F_\lambda$ of the AD is
the superimposition of the thermal spectra from concentric cylinder rings with
different radii
\begin{equation}\label{eqn:F_lambda}
  F_\lambda \propto \ R_{\rm S}^2 \ \int_{r_{\rm in}}^{r_{\rm out}} B_\lambda[T(r')]\ r'\,dr'
\end{equation}
where $r \equiv R/R_{\rm S}$ is the normalised radius in units of the Schwarzschild radius  $R_{\rm S}$,
$r_{\rm in}$ is the radius of the innermost stable circular orbit, and $r_{\rm out}$ is the outermost
radius of the AD. Assuming that the viscously dissipated gravitational energy of the in-falling matter is radiated away locally
by a black-body of the local effective temperature, the radial temperature profile of the AD is 
$T = 5.1 \,\cdot10^5 \ \dot{M}^{1/4} M_8^{-1/2} \, r^{-3/4} \, [1-(r/r_{\rm in})^{-1/2}]^{1/4}$
(Shakura \& Sunyaev \cite{Shakura73}) where
$M_8 = M/10^8 M_\odot$, $\dot{M}$ in $M_\odot$\,yr$^{-1}$, $T$ and $\tilde{T}$ in K.
Replacing $r$ by $s \equiv r/r_{\rm in}$ leads to the alternative expression
\begin{equation}\label{eqn:T1}
T         =  T^\ast \, s^{-3/4} \, \Big[ 1-s^{-1/2}\Big]^{1/4}
\quad\mbox{with}\quad
T^\ast = 2.2\cdot10^5\,\mbox{K} \ \Bigg( \frac{\dot{M}}{M_8^2} \Bigg) ^{1/4}
\end{equation}
(e.g., Pereyra et al. \cite{Pereyra06}).
The spectral shape of the continuum is completely determined by $T^\ast$ and $r_{\rm in}$.

The left panel of Fig.\,\ref{fig:AD_spectra} shows MTBB model spectra for $r_{\rm in} = 3$ (Schwarzschild BH)
and $T^{\ast} =  6\cdot 10^4\,\mbox{K}$ to $11\cdot 10^4$\,K (bottom to top, in steps of $10^4$ K).
The outer edge is set to $r_{\rm out} = 10^3$, the distance where the Toomre stability parameter of the disk
falls below unity under standard assumptions (Goodman \cite{Goodman03}). The exact value of $r_{\rm out}$ does
not have a significant influence on the results. Over-plotted in Fig.\,\ref{fig:AD_spectra}
are those model curves for $r_{\rm in} = 1$ (maximum rotating Kerr BH) that fit the $r_{\rm in} = 3$ model at
longer wavelengths. We do not want to confront specific AD models but just to illustrate roughly the combined  effect of
more realistic AD models (Hubeny et al. \cite{Hubeny00}, \cite{Hubeny01}).
We estimated a simple correction function by comparing the spectra
from the MTBB and the elaborated models of Hubeny et al. (\cite{Hubeny01}) for $M_8 = 1$ 
and we applied this correction to the spectra from all MTBB models. Though oversimplified, this naive treatment
may illustrate some general trends. The spectra from the corrected
MTBB (CMTBB) models (right panel of Fig.\,\ref{fig:AD_spectra}) have lower fluxes at long wavelengths and
higher fluxes at short wavelengths. (For $M_8 =1$ the transition is in the EUV at $\lambda \approx 200$\AA.)
For wavelengths longwards of Ly$\alpha$, the corrected spectrum is slightly flattened and a weak Balmer edge
is predicted. The Balmer edge becomes less pronounced for higher masses ($M_8 > 1$). Comptonisation has
only little effect when $M_8 > 1$ (Hubeny et al. \cite{Hubeny01}).

The comparison between the model spectra and the observed composites must be restricted to those spectral
windows where the contamination of the observed flux by emission lines is negligible.
In Fig.\,\ref{fig:AD_spectra}, five pseudo-continuum windows from Tsuzuki et al. (\cite{Tsuzuki06}) and
Sameshima et al. (\cite{Sameshima11}) are indicated. As mentioned above, host galaxy contamination can become
significant at $\lambda \ga 4000$\AA, dependent on the sample selection criteria. Taking these effects into account,
the agreement between the observed and the model spectra is very good for wavelengths longwards of the Ly$\alpha$ line.
The mean relative deviation between the SDSS composite spectrum and the best-fitting model curves is only a few percent.

%%%%%%%%%%%%%%%%%%%%%%%%%%%%%%%%%%%%%%%%%%%%%%%%%%%%%%%%%%%%%%%%%%%%%%%%%%%%%%%%%%%%%%%%%%%%%%%%%%%%%%%%%%%%%%%%%%
%
%
\section{Quasar sample and basic data}\label{sec:sample}
%
%
%%%%%%%%%%%%%%%%%%%%%%%%%%%%%%%%%%%%%%%%%%%%%%%%%%%%%%%%%%%%%%%%%%%%%%%%%%%%%%%%%%%%%%%%%%%%%%%%%%%%%%%%%%%%%%%%%%

%-------------------------------------------------------------------------------------------------------------------
%
\subsection{Black hole masses, luminosities, Eddington ratios}\label{sec:BH_mass}
%
%-------------------------------------------------------------------------------------------------------------------

The spectroscopic quasar catalogue from the SDSS DR7 (Schneider et al. \cite{Schneider10})
contains 105\,783 bona fide quasars with absolute i-band magnitudes $M_i < -22.0$.
The computation of BH mass estimates for such a large sample of quasars has become possible
by the availability of scaling relations (see e.g., Vestergaard \cite{Vestergaard11}).
The scaling relations are built on the assumptions that (1) the gas in the broad line region (BLR) is virialised,
(2) the width of the broad emission line is a proxy for the virial velocity, and (3) the distance of the
BLR from the central source scales with the continuum luminosity $\lambda L_\lambda$.
This approach enables estimating BH masses of quasars from single-epoch spectra.

Shen et al. (\cite{Shen11}) compiled a catalogue of properties of all quasars from the SDSS DR7,
including particularly the emission line widths of H$\beta$, \ion{Mg}{ii}, and \ion{C}{iv},
the continuum flux around these lines, and virial black hole masses $M$. The mass is computed from various
calibrations of the scaling relations. Shen et al. give also a fiducial BH mass derived
either from Vestergaard \& Peterson's (\cite{Vestergaard06}) calibrations for H$\beta$ and \ion{C}{iv}
or from their own calibration for \ion{Mg}{ii}.
In addition, $L_{\rm bol}$, $\varepsilon$, and many other parameters are provided. 
The bolometric luminosity was computed from the monochromatic
luminosity in a continuum window at 1350\AA, 3000\AA, or 5100\AA, respectively, using a
constant bolometric correction. In the following, we identify the BH mass with
the fiducial mass from Shen et al. (\cite{Shen11}).
Figure\,2 shows the distributions of the SDSS quasars in the $L_{\rm bol} - M - z$ space and in the
$\varepsilon - M - z$ space.

% Fig. 2a,b
\begin{figure}[htbp]
\includegraphics[width=9cm]{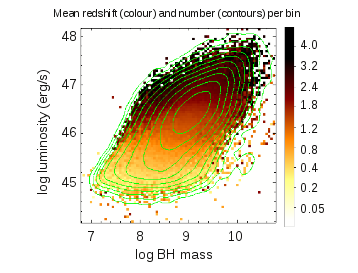}
\includegraphics[width=9cm]{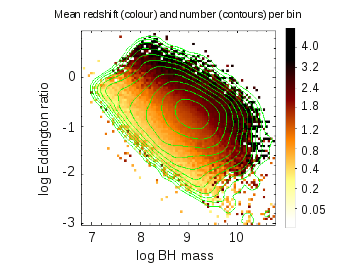}\\
\caption{Bolometric luminosity $L_{\rm bol}$ (top) and Eddington ratio $\varepsilon$
(bottom) versus BH mass $M$ ($M_\odot$) for the more than $10^5$ quasars from Shen et al. (\cite{Shen11}). The
colour coding indicates the redshift $z$. The number distribution of the quasars is marked by the
contour curves.}
\label{fig:stellar-cont}
\end{figure}

%-------------------------------------------------------------------------------------------------------------------
%
\subsection{Quasar sample}
%
%-------------------------------------------------------------------------------------------------------------------

We started with the quasar sample from Paper 1.
It was constructed from the inspection and analysis of the SDSS spectra of all those $2\,10^4$ objects
in S82 that were classified as quasars or quasar candidates
by the spectroscopic pipeline of the SDSS.  We selected 9855 quasars.
The variability data for the quasars were computed from the LMCC  (Bramich et al. \cite{Bramich08}).
The cross-correlation of the quasar list with the LMCC yields a reduced sample of 8744 quasars
because the LMCC covers not the whole S82. The redshifts cover the range $z \approx 0.1 - 5$ with the
maximum between $z \approx 1$ and 2. There was no absolute magnitude cut-off applied.
About 90\% of the quasars were identified in the SDSS DR7 quasar catalogue
(Schneider et al. \cite{Schneider10}). The majority of the remaining 10\% are low-$z$,
low-luminosity AGNs. The composite spectrum from our sample is in good agreement with the SDSS
quasar composite from Vanden Berk et al. (\cite{VandenBerk01}) in the UV but shows a stronger
contamination by the host galaxies of the low-$z$ quasars in the optical (Fig.\,\ref{fig:AD_spectra}).
For a more detailed description of the S82 quasar sample we refer to Paper 1.
We matched the variability catalogue from Paper 1 with the Shen et al. (\cite{Shen11}) catalogue and
identified 7990 quasars in both catalogues. 22 quasars have discrepant redshift values and
were simply excluded in the present study.

% Fig. 3
\begin{figure}[bhtp]
\includegraphics[width=9cm]{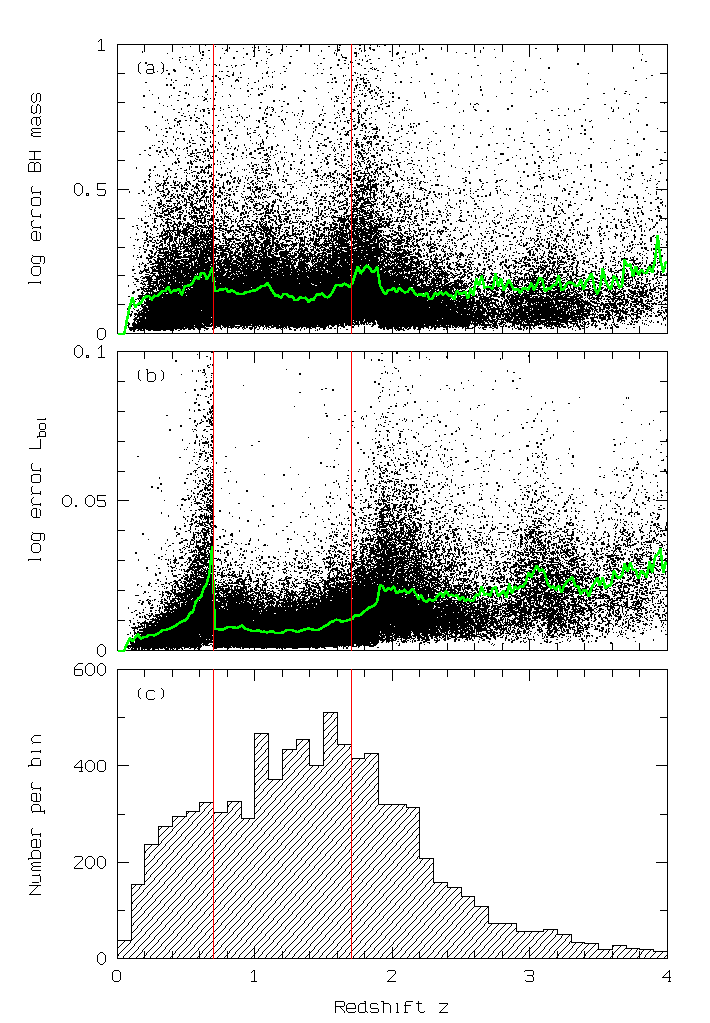}\\
\caption{
Errors of $M$ (a) and $L_{\rm bol}$ (b) as a function of redshift $z$ for the $10^5$ quasars
from Shen et al. (\cite{Shen11}). Polygon: mean values in narrow $z$ bins. Bottom (c): $z$ distribution
of the quasars from the variability sample from Paper 1.
Vertical lines: $z$ interval of the final quasar sample. For reasons of clarity, we refrain from plotting
very large errors in (a) and (b).
}
\label{fig:mass_error}
\end{figure}
%Weiss/Shen+/plot_errors_mass.prg

We aim at small uncertainties in the accretion parameters. The estimated accretion rate 
(Sect.\,\ref{sec:acc_rate}) is sensitive to the optical luminosity
$L_{\rm opt}$ and to the assumed mass. $L_{\rm opt}$ can be measured directly only at low $z$.
However, for the low-$z$ quasars the observed variability may be significantly diluted by
the host contribution (see Sect.\,\ref{sec:continuum}). Therefore, we consider higher-$z$ quasars where
$L_{\rm opt}$ has to be extrapolated from a monochromatic luminosity measured in the UV. The
extrapolation becomes more uncertain, of course, for too high redshifts. 

Figure \ref{fig:mass_error} shows the error of $M$ and $L_{\rm bol}$, respectively,
from the Shen et al. catalogue as a function of redshift. The fiducial mass uncertainty was
propagated from the measurement uncertainties of the continuum luminosities and the line widths,
but does neither include the statistical uncertainty ($\ga 0.3-0.4$\,dex) from the
calibration of the scaling relations, nor the systematic effects.
Because it is not clear whether the masses derived from the different lines (H$\beta$, \ion{Mg}{ii},
\ion{C}{iv}) are free from systematic effects, we decided to use only those quasars where the fiducial
masses were estimated from the \ion{Mg}{ii} line profile (i.e., $z \la 2$).
The formal BH mass errors as given by Shen et al. (\cite{Shen11})
depend on $z$ and increase significantly at those redshifts where the emission line used for
estimating the fiducial mass approaches an edge of the SDSS spectral window (Fig. \ref{fig:mass_error}a).
A similar effect is present also for the luminosity errors, namely when the continuum window where
the monochromatic luminosity is measured comes close to an edge
(Fig. \ref{fig:mass_error}b). The luminosity errors are smallest for $z = 0.7 - 1.7$ where the
mass errors are also comparatively small. We restricted the final quasar sample therefore to
the 3916 quasars in that redshift interval. This interval is close to
the maximum of the $z$ distribution of the SDSS quasars (Fig. \ref{fig:mass_error}c).

%-------------------------------------------------------------------------------------------------------------------
%
\subsection{Accretion rates and radiative efficiencies}\label{sec:acc_rate}
%
%-------------------------------------------------------------------------------------------------------------------

\subsubsection{Scaling relations}

% Fig. 4 a-h
\begin{figure*}[bhtp]
\begin{tabbing}
\includegraphics[viewport=40 -20 554 780,angle=270,width=8.4cm,clip]{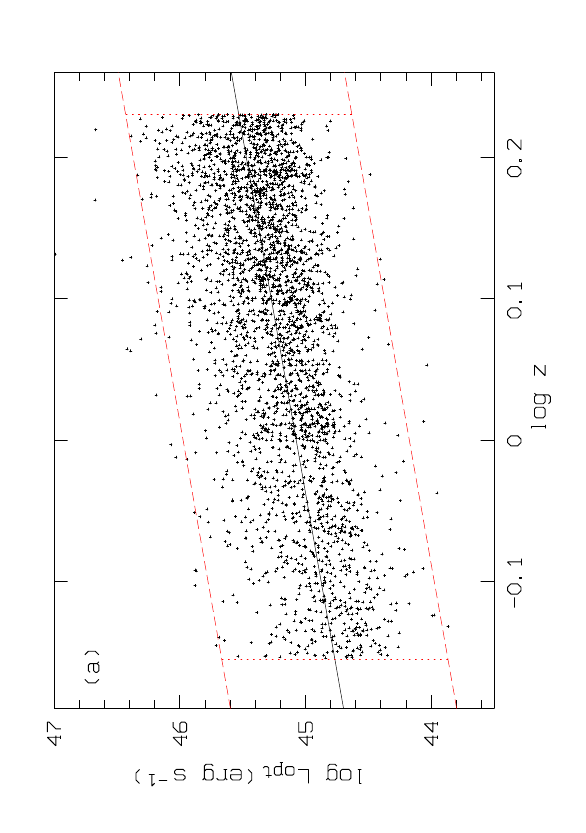}\hfill
\includegraphics[viewport=40 -30 554 770,angle=270,width=8.4cm,clip]{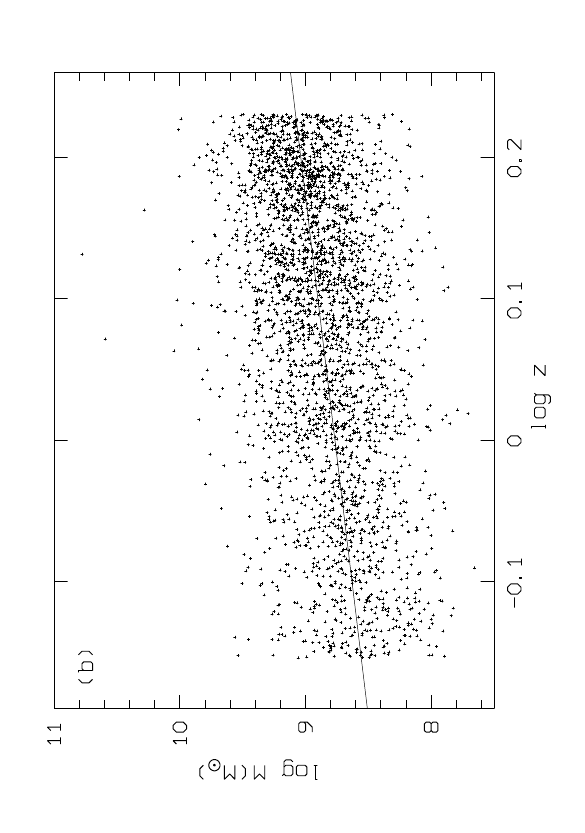}\hfill \\
\includegraphics[viewport=40 -20 554 780,angle=270,width=8.4cm,clip]{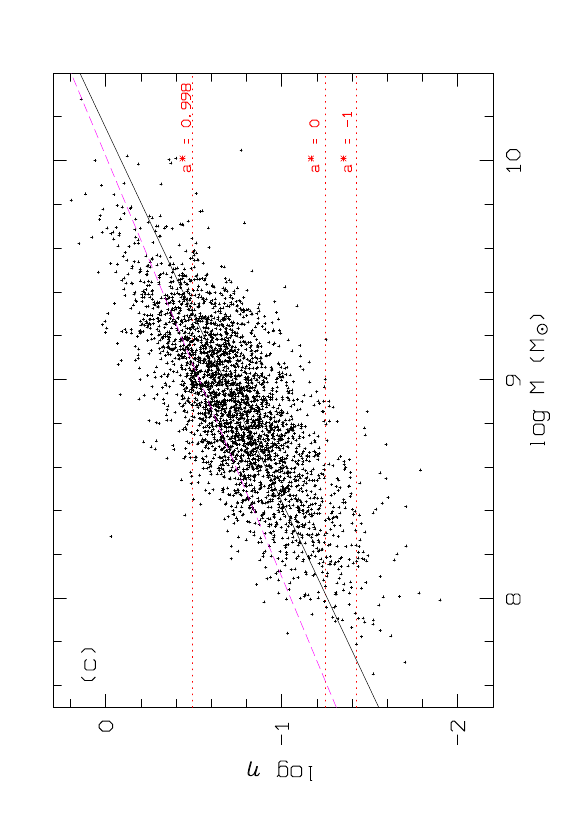}\hfill
\includegraphics[viewport=40 -30 554 770,angle=270,width=8.4cm,clip]{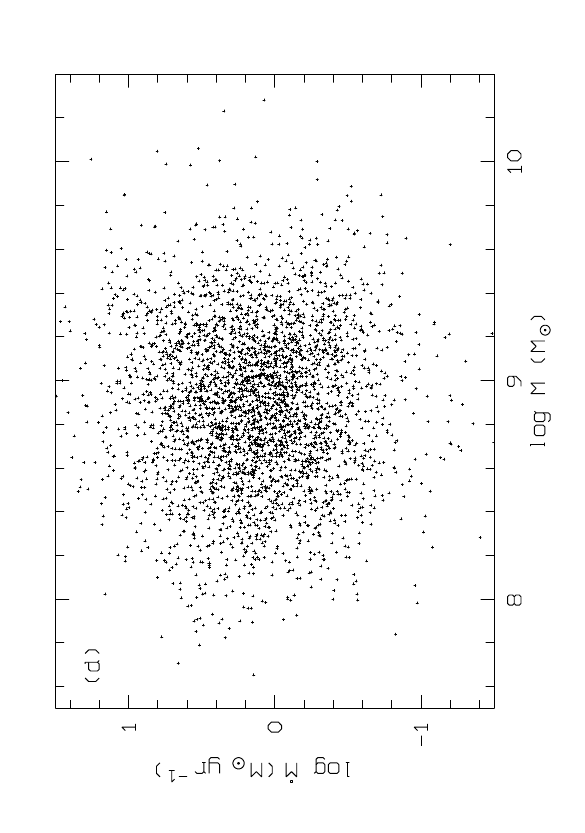}\hfill \\
\includegraphics[viewport=40 -20 554 780,angle=270,width=8.4cm,clip]{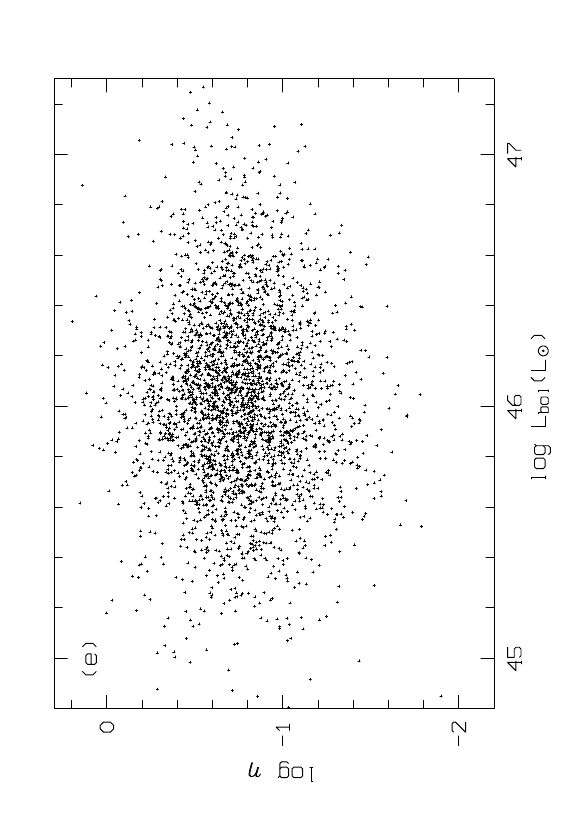}\hfill
\includegraphics[viewport=40 -30 554 770,angle=270,width=8.4cm,clip]{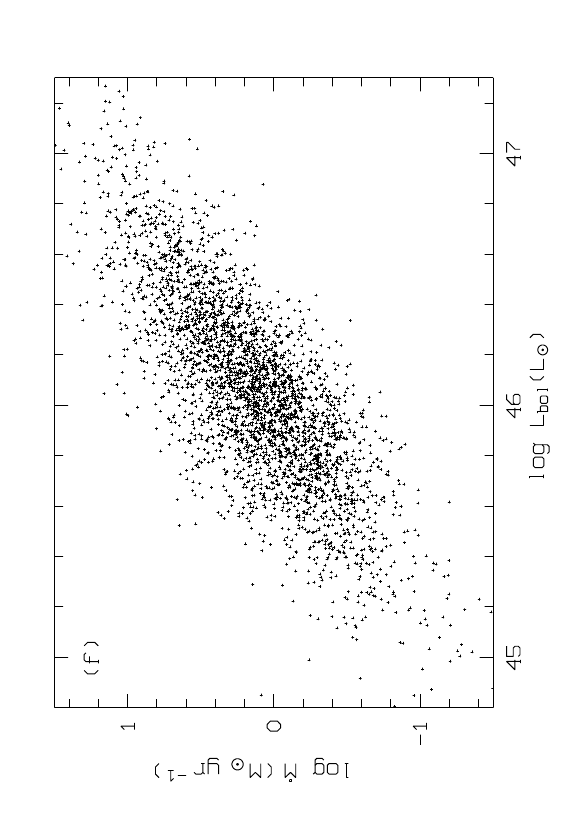}\hfill \\
\includegraphics[viewport=40 -20 554 780,angle=270,width=8.4cm,clip]{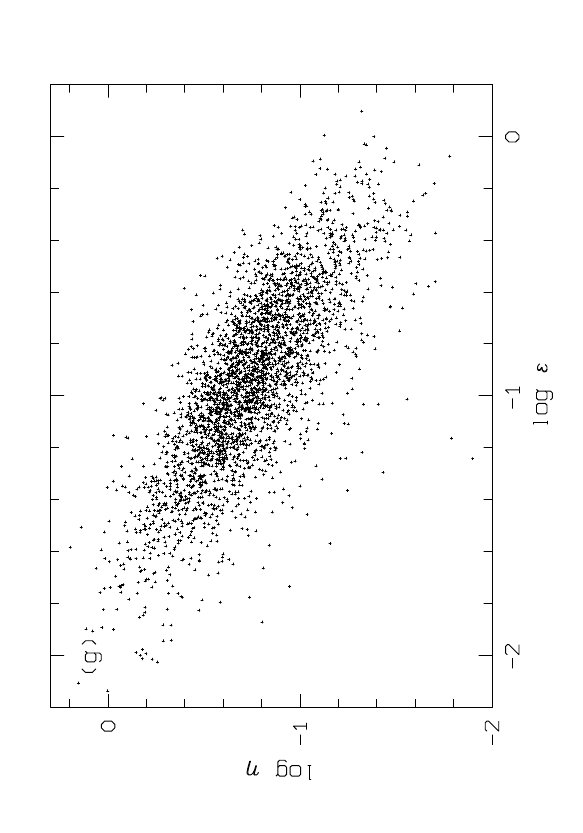}\hfill
\includegraphics[viewport=40 -30 554 770,angle=270,width=8.4cm,clip]{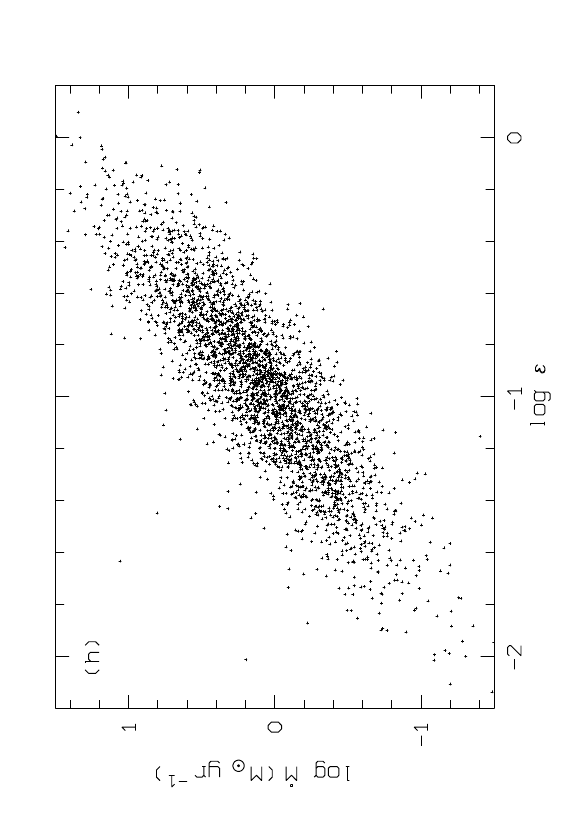}\hfill
\end{tabbing}
\caption{Properties of the sample of SDSS quasars.
Panels (a) and (b): Optical luminosity (a) and mass (b) as function of redshift (solid lines:
linear regression curves, other lines in (a): limits for our simulated sample).
Other panels: radiative efficiency $\eta$ and mass accretion rate $\dot{M}$
as function of mass (c,d), bolometric luminosity (e,f) and Eddington ratio $\varepsilon$ (g,h).
Thin diagonal lines in panel (c): linear regression curves from DL11 (dashed) and
from our SDSS sample (solid); horizontal dotted lines: Novikov-Thorne efficiencies
for three critical values of the BH spin parameter $a^\ast$.
See the text for a more detailed description.
}
\label{fig:eta_M}
\end{figure*}
% /Weiss/Shen+11/plot_Lopt_z.prg
% /Weiss/Shen+11/plot_M_z.prg
% /Weiss/Shen+11/plot_eta_M_S82_3.prg
% /Weiss/Shen+11/plot_dM_M.prg
% /Weiss/Shen+11/plot_eta_L.prg
% /Weiss/Shen+11/plot_dM_L.prg
% /Weiss/Shen+11/plot_eta_epsilon.prg
% /Weiss/Shen+11/plot_dM_epsilon.prg

The rate of the mass accretion onto the BH is directly related to the total energy
radiated by the quasar:
\begin{equation}\label{eqn:rad_eff}
 L_{\rm bol} = \eta \, \dot{M} \,c^2,
\end{equation}
where $\eta$ is determined by the smallest stable orbit behind which matter
is expected to fall directly into the BH without emitting radiation, i.e., it depends on the spin of the BH.
$L_{\rm bol}$ can be taken as a proxy for the accretion rate. However, both the radiative efficiency
$\eta$ and the bolometric correction suffer from substantial uncertainties.
Estimates of the average radiative efficiency using different methods
yield $\bar{\eta} \ga 0.1$
(Soltan \cite{Soltan82};
Elvis et al. \cite{Elvis02};
Yu \& Tremaine \cite{Yu02};
Wang et al. \cite{Wang06};
Davis \& Laor \cite{Davis11}; hereafter DL11).
Individual values of $\eta$ may not only scatter over about one order of magnitude
(depending on the accretion history from chaotic to coherent accretion; see
Dotti et al. \cite{Dotti13}) but may also depend on $M$. A correlation between
spin and mass was predicted by numerical simulations of the BH growth
(Dotti et al. \cite{Dotti13}) and seems to be present in different
quasar samples
(DL11;
Laor \& Davis \cite{Laor11a};
Li et al. \cite{Li12};
Raimundo et al. \cite{Raimundo12};
Chelouche \cite{Chelouche13};
see also below).
At highest temperatures, the $T$ profile is determined by $r_{\rm in}$ (Eq. \ref{eqn:T1}).
Hence, the uncertainty of the bolometric correction is mainly produced by the uncertain fraction
of energy radiated in the X-ray and EUV domains from the innermost part of the AD.

Estimating the accretion rate of a steady-state AD  from the optical luminosity
is far less affected by the uncertain innermost part of the disk.
For $r \gg r_{\rm in}$ the temperature profile (Eq.\,\ref{eqn:T1}) becomes
$T \approx T^\ast\,r^{-3/4}$  and the integral in Eq.\,(\ref{eqn:F_lambda})
can be expressed by the simple relationship  between luminosity, accretion rate, and
black hole mass
$ \dot{M} = 2.0\,L_{\rm opt, 45}^{1.5}\, M_8^{-1}$
(DL11\footnote{There was a numerical
error in the corresponding  equations (5) and (7) in DL11. The
correct numerical factors are given  in a footnote by Laor \& Davis (\cite{Laor11b}).})
for $\lambda = 4681$\AA\ and  a typical disk inclination of $\cos\,i = 0.8$ where
$L_{\rm opt,45} = \lambda L_\lambda$ is in units of $10^{45}\,\mbox{erg}\,\mbox{s}^{-1}$.
DL11 demonstrated that more sophisticated models provide similar
relations. In particular, they found
$\dot{M} = 3.5\,L_{\rm opt, 45}^{1.5}\, M_8^{-0.89}$ (their Eq. 8)
from low-$z$ Palomar-Green quasars where $M$ was derived from the bulge stellar
velocity distribution. DL11 suggested that their simple fitting
relations ``could be used in place of a detailed model fitting method for future work''.

We applied Eq. (8) from DL11 to the quasars in our sample. However, because the optical luminosity
is not directly accessible for the majority of the SDSS quasars we estimated  $L_{\rm opt}$
by the extrapolation of the monochromatic luminosity at 3000\AA, $L_{3000}$, from Shen et al.
(\cite{Shen11}) adopting the power law from Eq.\,(\ref{eqn:power-law}):
\begin{equation}\label{eqn:AR_DL}
 \dot{M} = 3.5\cdot 0.64^{-1.5\cdot(1+\alpha_\lambda)}\, L_{3000, 45}^{1.5}\, M_8^{-0.89}.
\end{equation}
We derived the spectral slope $\alpha_\lambda$  for each quasar individually
by fitting a power law to the foreground extinction-corrected SDSS spectrum in
the pseudo-continuum windows. These individual continuum slopes were used for all those
quasars having spectra with $S/N>3$ in at least three windows. For quasars with noisier spectra we simply
adopted the mean slope $\alpha_\lambda = -1.52$ from the composite spectrum of the parent sample (Paper 1).
The overwhelming majority of the quasars in our sample have accretion rates within $\pm 1$ dex 
around a mean value of about $1 M_\odot$\,yr$^{-1}$.

DL11 estimated $\dot{M}$ by fitting model spectra to the individual optical spectra.
The bolometric luminosity could be estimated from the broadband spectral energy distribution
because the quasars in their sample have ample coverage from optical to X-ray,
with some uncertainty in the EUV. DL11 used $L_{\rm bol}$ and $\dot{M}$ to compute the
radiative efficiency $\eta$ from Eq.\,(\ref{eqn:rad_eff}).

Here, we applied Eqs.\,(\ref{eqn:AR_DL}) and (\ref{eqn:rad_eff})
in combination with the assumption of a constant bolometric correction
$K = L_{\rm bol}/L_{\rm opt}$ to estimate $\eta$ from the optical luminosity as
\begin{equation}\label{eqn:rad_eff_2}
 \eta = 6.25\cdot 10^{-3}  M_8^{0.89} L_{\rm opt,45}^{-0.5}
\end{equation}
where we assumed $K=6.3$ (mean value in the final sample) and an AD inclination $\cos\,i = 0.8$.
The assumption of a bolometric luminosity is supported by the
findings of a constant ratio $L_{\rm opt}/L_{\rm bol}$ by DL11 and an only weak dependence
of this ratio on $L_{\rm bol}$ by Marconi et al. (\cite{Marconi04}). The resulting mean radiative efficiency
is $\bar{\eta} = 0.22$. Most quasars ($76\%$) have values between those corresponding
to the Novikov-Thorne efficiencies $0.057 < \eta < 0.321$ for $0 < a^\ast < 0.998$. For 19\% of the
quasars the estimated $\eta$ is higher than the upper limit for $a^\ast = 0.998$ and a very small
fraction (0.5\%) has $\eta>1$. Given the uncertainties in $\dot{M}$ and $M$, these results can be considered
as broadly consistent with the Novikov-Thorne limits.

\subsubsection{Correlation diagrams}

% Fig. 5a-h
\begin{figure*}[bhtp]
\begin{tabbing}
\includegraphics[viewport=0 0 500 330,angle=0,width=8.5cm,clip]{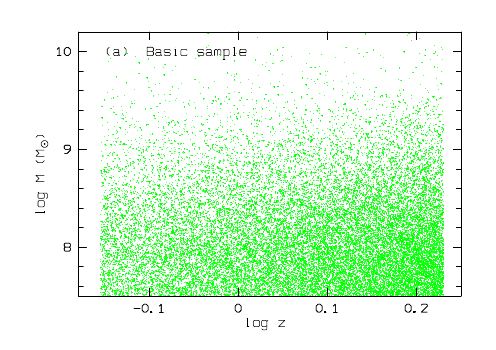}\hfill
\includegraphics[viewport=0 0 500 330,angle=0,width=8.5cm,clip]{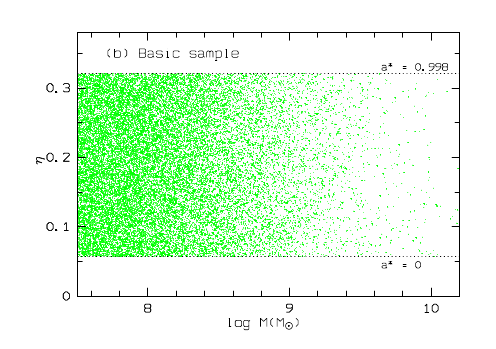}\hfill \\
\includegraphics[viewport=0 0 500 330,angle=0,width=8.5cm,clip]{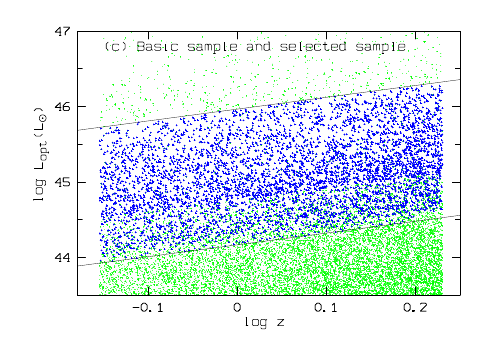}\hfill
\includegraphics[viewport=0 0 500 330,angle=0,width=8.5cm,clip]{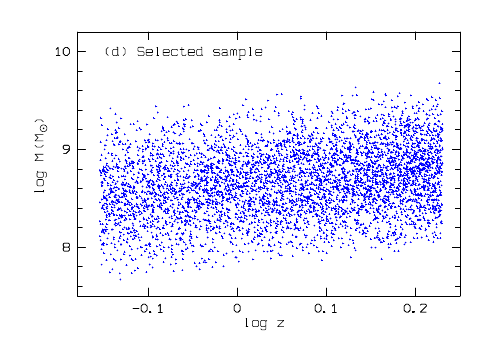}\hfill \\
\includegraphics[viewport=0 0 500 330,angle=0,width=8.5cm,clip]{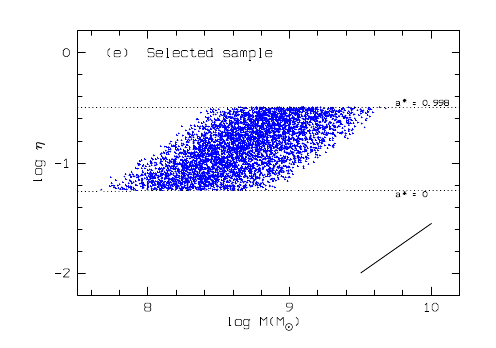}\hfill
\includegraphics[viewport=0 0 500 330,angle=0,width=8.5cm,clip]{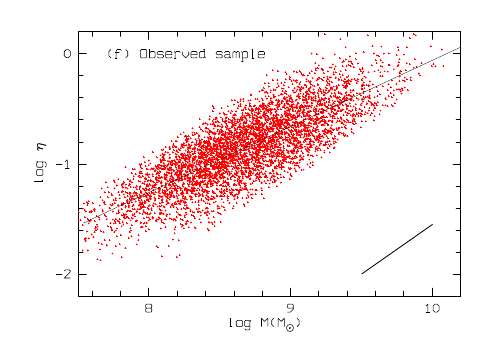}\hfill \\
\includegraphics[viewport=0 0 500 330,angle=0,width=8.5cm,clip]{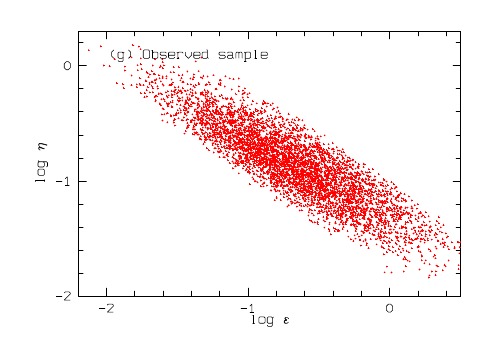}\hfill
\includegraphics[viewport=0 0 500 330,angle=0,width=8.5cm,clip]{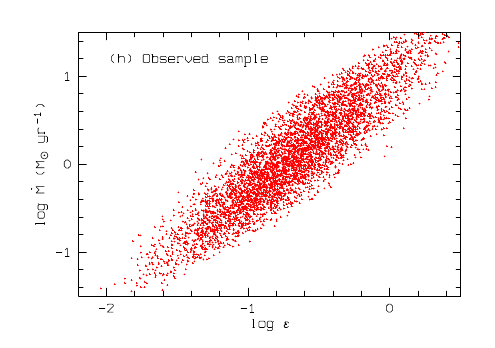}\hfill
\end{tabbing}
\caption{Simulated basic quasar sample (a-c; small green dots), selected sample (c-e; small blue crosses),
and simulated ``observed'' quasar sample (f-h; small red crosses). The selection is based solely on the
luminosity limits of the SDSS sample (c; see also Fig.\ref{fig:eta_M}a).
The thick diagonal lines in the lower right corner of panels e and f indicate the displacement for a 0.5 dex
shift in $M$. The ``observed'' sample is identical with the selected sample but mass errors were involved in the latter.
Panels (f) to (h) show correlations in the ``observed'' sample. Solid line in (f): linear regression curve.
See the text for a more detailed description.
}
\label{fig:eta_M_sim}
\end{figure*}
% /Weiss/Shen+11/simulations/plot_sim_M_z_all.prg
% /Weiss/Shen+11/simulations/plot_sim_eta_M_all.prg
% /Weiss/Shen+11/simulations/plot_sim_L_z.prg
% /Weiss/Shen+11/simulations/plot_sim_M_z_selected.prg
% /Weiss/Shen+11/simulations/plot_sim_eta_M_selected.prg
% /Weiss/Shen+11/simulations/plot_sim_eta_M_with_errors.p
% /Weiss/Shen+11/simulations/plot_sim_eta_eps.prg
% /Weiss/Shen+11/simulations/plot_sim_Mdot_eps.prg

\noindent{\it (a) \ Observed correlations}\\

Figure\,\ref{fig:eta_M} displays eight correlation diagrams for the quasars in our final SDSS quasar sample.
These diagrams will be briefly discussed in this paragraph and will be analysed further in paragraph (b) below.

Figure\,\ref{fig:eta_M}a shows $L_{\rm opt}$ versus redshift. 
As expected for a flux-limited sample, there is a clear trend of $L_{\rm opt}$ to increase with $z$. 
The diagram illustrates the major selection effects inherent in our quasar sample: the strictly 
defined $z$ interval (vertical dotted lines)
and the $z$-dependent $L_{\rm opt}$ interval (diagonal dashed lines; see the discussion in paragraph (b) below).
There is also a tendency for the masses to increase with $z$ (Fig.\,\ref{fig:eta_M}b). 
The Eddington ratio $\varepsilon$ (not shown) has only a weak trend with $z$. 
Linear regression yields $\log \varepsilon \propto -0.42 \log M$ and 
$\log \varepsilon \propto 0.43 \log L_{\rm bol}$.

Figure\,\ref{fig:eta_M}c reveals a remarkable trend towards higher radiative efficiencies at higher masses.
The best-fit relation $\log \eta = -2.25+0.57 \log M_8$ from the linear regression roughly agrees with
$\log \eta = -2.05+0.52 \log M_8$ derived by DL11 (see also Laor \& Davis \cite{Laor11a}) from their sample of 80
low-$z$ Palomar-Green quasars. There is, on the other hand, no significant mass dependence of $\dot{M}$ (Fig.\,\ref{fig:eta_M}d).

Figures\,\ref{fig:eta_M}e,f indicate that the situation is opposite for the dependences on the luminosity:
$\eta$ is uncorrelated with $L_{\rm bol}$ while $\dot{M}$ and  $L_{\rm bol}$ are strongly correlated.
Finally, there is a strong anti-correlation of $\varepsilon$ and $\eta$ (Fig.\,\ref{fig:eta_M}g), where we find
$\varepsilon \propto \eta^{-0.89}$ as the best-fit relation, and a tight correlation
($\log \varepsilon \propto 0.62 \log \dot{M}$) between $\varepsilon$ and $\dot{M}$
(Fig.\,\ref{fig:eta_M}h), i.e., the Eddington ratio is a good proxy for the accretion rate.

All relations shown in Figs.\,\ref{fig:eta_M}c-h can be understood from 
Eqs.\,(\ref{eqn:AR_DL}) and (\ref{eqn:rad_eff_2}) in combination with the observed 
$\varepsilon-L_{\rm bol}$ and $\varepsilon-M$ relations 
and assuming a constant bolometric correction, i.e., $L_{\rm opt} \propto L_{\rm bol}$. 
For example, $\varepsilon \propto M^{-0.42}$ means
$L_{\rm bol} \propto M^{0.58}$.  Replacing the luminosity in Eq.\,(\ref{eqn:rad_eff_2})  
yields $\eta \propto M^{0.60}$, very close to the observed relation (Fig.\,\ref{fig:eta_M}c). 
On the other hand, it is easily seen that $\eta$ should be independent of $L_{\rm bol}$  
(Fig.\,\ref{fig:eta_M}c): the observed relation $\varepsilon \propto L_{\rm bol}^{0.43}$ 
means $M \propto L_{\rm bol}^{0.57}$ and thus
$\eta \propto L_{\rm bol}^{0.507} L_{\rm bol}^{-0.5} \approx 1$.
Further, replacing the luminosity in Eq.\,(\ref{eqn:AR_DL}) shows that $\dot{M}$ is nearly 
independent of $M$ (Fig.\,\ref{fig:eta_M}d). In a similar way
one finds $\eta \propto \varepsilon^{-0.89}M^{0.22} \propto \varepsilon^{-1}$ (Fig.\,\ref{fig:eta_M}g)
and $\dot{M} \propto \varepsilon$ (Fig.\,\ref{fig:eta_M}h).
However, this internal consistency of the observed correlations does not explain 
the underlying reason for the correlations, particularly for the $\eta-M$ relation.\\

\noindent{\it (b) \ Simulations}\\

The accretion theory explains different radiation efficiencies as an effect of different
black hole spins. A systematic trend of $\eta$ with $M$, as found by DL11 and apparently
confirmed by our larger sample of higher-$z$ quasars, could provide an important constraint on the evolution of
SMBHs (Dotti et al. \cite{Dotti13}) if it could be unambiguously shown to be not
produced by selection effects and uncertainties in the input data. DL11 discussed the effects 
of the errors in $M$ and disfavoured mass measurement errors alone as an explanation for the observed 
correlation. Raimundo et al. (\cite{Raimundo12}) argued that the distribution of $\eta$ in the DL11 
sample is shaped by the selection criteria of the sample in combination with an almost constant 
bolometric correction. To verify this argument, they used a simulated quasar sample.

Here we followed a similar approach to understand the correlation diagrams in Fig.\,\ref{fig:eta_M}.
The purpose of this exercise is to investigate whether properties that are 
intrinsically uncorrelated in the underlying basic (parent) sample display apparent 
correlations in the observed sample as a consequence of selection effects.
We simulated a quasar sample with a  $z$ distribution similar to that of our SDSS quasar sample 
(Fig.\,\ref{fig:mass_error}c) and with a  ``reasonable mass'' distribution.
Modelling a realistic SMBH mass function is clearly beyond the scope of the present paper. 
Instead we simply assumed a $z$-independent one-sided Gaussian to
account for the decrease of the mass distribution at high masses (Fig.\,\ref{fig:eta_M_sim}a). 
This is our underlying basic sample. Next, we assigned a randomly chosen value of $\eta$ to each quasar, 
where $\eta$ is uniformly distributed between the Novikov-Thorne limits 0.057 and 0.321
for $a^\ast = 0$ and $a^\ast = 0.998$, respectively (Fig.\,\ref{fig:eta_M_sim}b).
There is no correlation of $\eta$ with $M$ in the basic sample. Now, with $\eta$ and $M$ given, 
we used Eqs.\,(\ref{eqn:rad_eff}) and (\ref{eqn:AR_DL}) to compute the optical luminosity 
$L_{\rm opt}$ for each quasar adopting a constant bolometric correction.

A general selection effect in real quasar samples is produced by the flux limitation of the survey.
The flux limit produces a lower luminosity cut that increases with redshift. 
About half of the objects in the Shen et al. (\cite{Shen11}) catalogue were selected uniformly 
using the final quasar target selection algorithm (Richards et al. \cite{Richards02}), the 
other half were selected via various algorithms. That is, the distribution in the $L-z$ 
plane shows a sharp lower $L(z)$ limit only for the uniformly-selected subsample whereas the other 
quasars populate also the area below that lower limit (see e.g., Fig.\,1 of Shen et al. \cite{Shen11}). 
The lower luminosity limit is thus not strictly defined in our sample.
The upper luminosity limit, on the other side, is set mainly by the quasar luminosity function and
is also not well defined. Therefore we simply used the selection area indicated in
Fig.\,\ref{fig:eta_M}a for our SDSS sample and selected all quasars from the basic sample within 
this area. To account for the inhomogeneity of the SDSS quasar catalogue, and thus for the incompleteness
of our SDSS sample near the lower luminosity limit, the lower luminosity part of the selected (simulated)
sample has been randomly thinned out in a somewhat arbitrary way.
Fig.\,\ref{fig:eta_M_sim}c  illustrates the selection. Fig.\,\ref{fig:eta_M_sim}d shows the 
distribution of the  selected sample on the $M-z$ plane.

The selected quasars show a rhomboidal distribution in the $\eta-M$ plane (Fig.\,\ref{fig:eta_M_sim}e).
That means there is a tendency towards higher efficiencies for higher masses, even though $\eta$ and $M$
are not correlated in the underlying basic sample. The shape of this distribution can be easily
explained:  The lower and upper edges are set by the limits of the SMBH spin.
At a fixed value of $\eta$, the black hole masses are distributed over the range 
$k \eta^{1.12} (L_{\rm opt,low}^{0.56} \ldots L_{\rm opt, up}^{0.56})$ where 
$L_{\rm opt,low}$ and $L_{\rm opt, up}$, respectively, are the lower and higher luminosity limits
and $k$ is a constant value. Hence, the limits on the left and on the right hand side, respectively, 
reflect the luminosity limits of the survey. 

The observed quasar sample is subject to uncertainties of the mass estimation. The direction 
of the displacement in the $\eta-M$ plane is indicated by the diagonal line in Fig.\,\ref{fig:eta_M_sim}e.
Mass errors have the effect to enhance the trend of increasing mean $\eta$ with $M$ 
seen in Fig.\,\ref{fig:eta_M_sim}e. From the uncertainties  given in the Shen et al. catalogue we
found a weakly $L_{\rm opt}$-dependent standard deviation
$\sigma(\log M) = 0.26$ - $0.03\cdot \log L_{\rm opt,45}$ 
for our SDSS quasar sample. We adopted this relation for simulating Gaussian distributed mass errors. 
Luminosity errors are much smaller and can be neglected.
The thus corrected data from the selected simulated sample constitute our ``observed'' 
(simulated) sample.  The distribution in the $\log \eta - \log M$ plane (Fig.\,\ref{fig:eta_M_sim}f)
is very similar to that of the SDSS quasar sample (Fig.\,\ref{fig:eta_M}c),
the linear regression yields a slope of 0.60, compared to 0.57 in the SDSS sample.

We conclude that the observed $\eta - M$ relation is mainly caused by the selection effects in 
combination with the mass errors and the scaling relation for $\dot{M}$.
A similar result was reported in a recent paper by Wu et at. (\cite{Wu13}) based on a mock sample
of quasars according to more realistic distributions of black hole masses and Eddington ratios.
For the sake of
completeness we note that also the other observed correlations are reproduced by the 
``observed'' (simulated) sample (compare Figs.\,\ref{fig:eta_M_sim}g and h with Figs.\,\ref{fig:eta_M}g and h).

%-------------------------------------------------------------------------------------------------------------------
%
\subsection{Variability estimator}\label{sec:var}
%
%-------------------------------------------------------------------------------------------------------------------

The variability estimator was described in Paper\,1. Here we briefly summarise only a
few facts that may be important in the context of the present paper and refer to Paper\,1 for more details.
We computed the rest-frame first-order structure function (SF)  $D(\tau_{\rm rf})$ for
each light curve in each band from the LMCC light curves.
The SF is a sort of running variance of the magnitudes as a function
of the (rest-frame) time-lag $\tau_{\rm rf}$, i.e., the time difference between two
arbitrary measurements. $D(\tau_{\rm rf})$ increases monotonically up to the characteristic
variability time-scale. As the maximum variability
time-scale of quasars is not well known, the time-lags must be long
in order to characterise the strength of the variability by $D$.
We defined the individual variability estimator $V_m$ of a quasar by
the arithmetic mean of all noise-corrected SF data points in the interval
$\tau_{\rm rf, max} \approx 1-2$\,yr. In other words, the quantity used to
describe the strength of the variability of a quasar is the variance of its magnitude differences
from all those pairs of two measurements that have rest-frame time-lags between 1 and 2 yr.
$V_m$ was computed for each of the five SDSS bands.\footnote{The index ``m'' indicates that the variance
is related to magnitudes.}

The dominant variability modes are associated with physical processes playing a role in the behaviour of disks.
These processes are often characterised by fundamental time-scales (e.g.,
Collier \& Peterson \cite{Collier01};
Frank \cite{Frank02})
such as the orbital time-scale $t_{\rm orb} = 1/\Omega$,
where $\Omega$ is the angular frequency, and the thermal time-scale $t_{\rm th} = t_{\rm orb}/\alpha$,
where $\alpha$ is is the viscosity parameter.
Using the radial temperature profile of the standard AD
in combination with Wien's displacement law, the radius dependence
of the time-scales can be transformed into a wavelength dependence. Furthermore, adopting
the relation
$\eta = 0.056\,M_8^{0.57}$
for the quasars in our sample (Sect.\,\ref{sec:acc_rate}),
the thermal time-scale (rest-frame) can be written as
\begin{displaymath}
t_{\rm th}  = 0.23 \ M_8^{0.21} \ \frac{\varepsilon_{0.1}^{1/2}}{\alpha}
               \ \Big(\frac{\lambda_{\rm obs,5000}}{1+z}\Big)^2 \ \mbox{yr}
\end{displaymath}
with $\varepsilon_{0.1} = \varepsilon/0.1$ and $\lambda_{\rm obs, 5000} = \lambda_{\rm obs}/5000$\AA,
where $\lambda_{\rm obs}$ is the wavelength in the observer frame.
Assuming $M_8=1-10$,
$\varepsilon_{0.1}=1, \lambda_{\rm obs,5000} = 1$, $z=1.5$, and
$\alpha \approx 0.03 - 0.2$ (e.g., Liu et al. \cite{Liu08}) we find
$t_{\rm th} \approx 0.2-2\,\mbox{yr} \la \tau_{\rm rf, max}$.
Hence, if quasar variability is dominated by thermal processes,
this estimate supports the assumption that the chosen time-lag interval is suitable to
derive a characteristic variability indicator.
(Needless to say that the viscosity parameter $\alpha$ is poorly known however.)
The viscous time-scale, i.e., the time-scale of the accretion flow to
radiate away the energy released by viscous dissipation, is $t_\nu \approx t_{\rm th} (r/h)^2$,
where $h$ is the scale height of the AD in units of $R_{\rm S}$.
For a thin AD with $h \ll r$ one has to expect therefore $t_\nu \gg \tau_{\rm rf, max}$.
From the analysis of the variability of SDSS quasars by means of a random walk model, a
characteristic variability time-scale of SDSS quasars of about $200$ days was derived (Kelly et al.
\cite{Kelly09}; MacLeod et al. \cite{MacLeod10}).

%%%%%%%%%%%%%%%%%%%%%%%%%%%%%%%%%%%%%%%%%%%%%%%%%%%%%%%%%%%%%%%%%%%%%%%%%%%%%%%%%%%%%%%%%%%%%%%%%%%%%%%%%%%%%%%%%%%%
%
%
\section{Slope-luminosity effects}\label{sec:slope-luminosity}
%
%
%%%%%%%%%%%%%%%%%%%%%%%%%%%%%%%%%%%%%%%%%%%%%%%%%%%%%%%%%%%%%%%%%%%%%%%%%%%%%%%%%%%%%%%%%%%%%%%%%%%%%%%%%%%%%%%%%%

%-------------------------------------------------------------------------------------------------------------------
%
\subsection{Slope-luminosity and slope-temperature relations}
%
%-------------------------------------------------------------------------------------------------------------------

\begin{figure}[bhtp]
\includegraphics[width=6.5cm,angle=270]{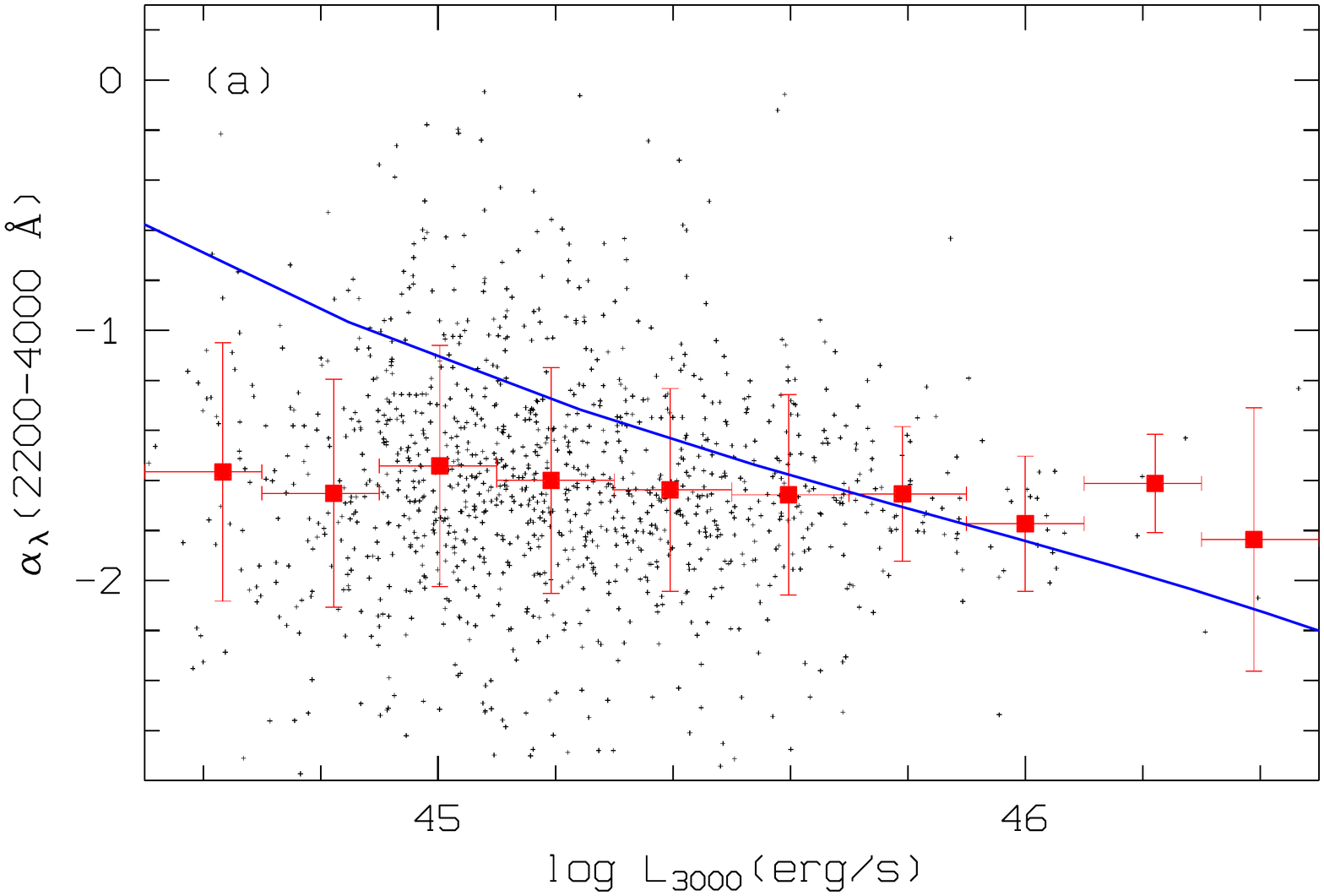}
\includegraphics[width=6.5cm,angle=270]{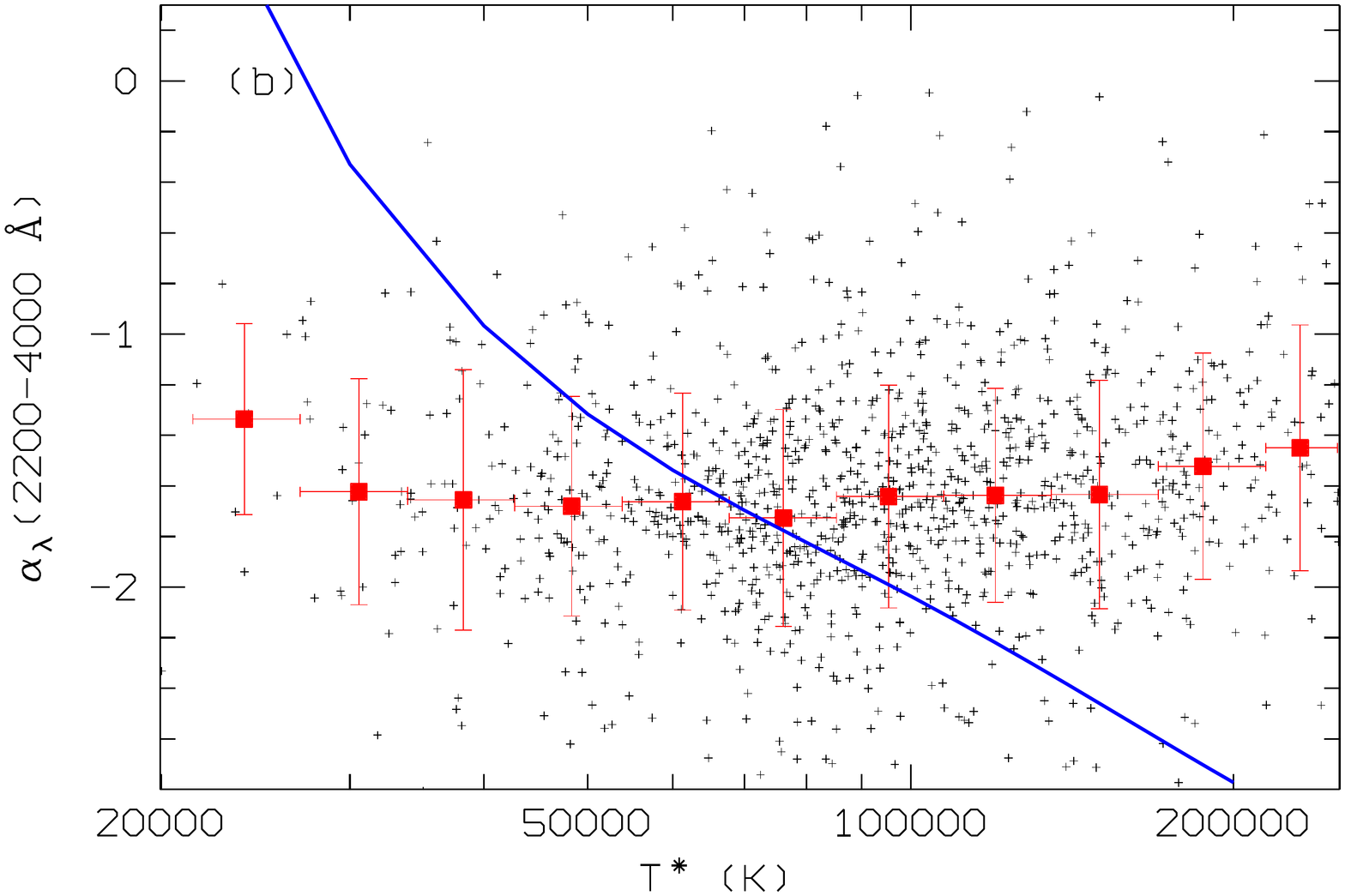}
\caption{
Spectral slope $\alpha_\lambda$ versus $\log\ L_{3000}$ (a) and $T^\ast$ (b),
respectively. Filled squares with vertical error bars (standard deviation) for median values of binned date; horizontal bars: bin width.
Solid curves: expected relations from the MTBB models.}
\label{fig:alpha_T}
\end{figure}

For the standard AD one expects a steeper spectral slope $\alpha_\lambda$ for more luminous quasars (Fig.\,\ref{fig:AD_spectra}).
A slope-luminosity effect has been known for a long time
(e.g., Mushotzky \& Wandel \cite{Mushotzky89}; Lawrence \cite{Lawrence05}; Davis et al. \cite{Davis07}).
In Fig.\,\ref{fig:alpha_T}a, $\alpha_\lambda$ is plotted versus $L_{3000}$ for those $1200$ quasars
where the continuum flux is measured in the windows at $2200$\AA\ and $4200$\AA\ (i.e., $0.7 \le z \le 1.2$)
with $S/N>5$ (see Sect.\,\ref{sec:acc_rate}). There is no significant trend in our data.
Davis et al. (\cite{Davis07}) measured the spectral slopes of two large samples of SDSS quasars at 2200-4000\AA\ and
1450-2200\AA, respectively. No clear variation with luminosity was found for
$\alpha_\lambda (1450-2200\AA)$ but was indicated for $\alpha_{\lambda} (2200-4000\AA)$ with steeper (bluer)
slopes at higher luminosity. This trend was found to be significant, though smaller than predicted by the model.
One scenario for differences of the $\alpha_{\lambda} (2200-4000\AA)$ values from Davis et al. and ours
could be that the measured flux is contaminated by the host galaxy and
that this contamination is stronger at 4200\AA\ (here) than at 4000\AA\ (Davis et al.). Consequently, our slopes would be
slightly flatter (redder). However, if the host contribution is independent of the quasar luminosity, the flattening
effect should be stronger at low luminosities, where our slopes tend to be slightly steeper (bluer) compared to Davis et al.
Another possible reason could be differences in the composition of the quasar samples.
The Davis et al. sample covers a slightly higher redshift interval ($z = 0.76-1.26$) and is about
three times larger than ours (because of the $S/N$ criterion in our sample). As a consequence, our sample covers only 1.5 dex
in luminosity, compared to 2 dex for the Davis et al. sample. Assuming a mean spectral slope $\alpha_\lambda = -1.5$, our
luminosity interval $L_{3000} \approx 44.6-46.1$\,erg\,s$^{-1}$ corresponds to
$L_{2200} \approx 44.8-46.3$\,erg\,s$^{-1}$ where the mean slopes $\alpha_\lambda (2200-4000\AA)$ in luminosity bins from
Davis et al. (their Fig. 3) changes from $-1.5$ to $-1.8$ with increasing $L_{2200}$ and to
$L_{4000} \approx 44.8-46.3$\,erg\,s$^{-1}$ where $\alpha_\lambda (2200-4000\AA)$ varies with increasing $L_{4000}$
from $-1.6$ to $-1.7$ (their Fig. 5).
These differences are clearly within our error bars and 
we argue therefore that Fig.\,\ref{fig:alpha_T}a is not contradictory to Davis et al.  (\cite{Davis07}).

% Fig. 6 a,b
% ../Shen+11/plot_alpha_L3000_with_model_paper.prg
% ../Shen+11/plot_alpha_T_ast_with_model_paper.prg 

The discrepancy between the observations and the model prediction becomes stronger
when the dependence of $\alpha_\lambda$ on the disk temperature is considered (Fig.\,\ref{fig:alpha_T}b).
The temperature parameter $T^\ast$ was computed from Eq.\,(\ref{eqn:T1}) with $\dot{M}$ from
Eq.\,(\ref{eqn:AR_DL}). At low temperatures the observed slopes show a trend to become slightly steeper
(bluer) with increasing disk temperature, but this trend inverts at $T^\ast \approx 40\,000$\,K.
From Wien's displacement law, one would naively expect that cooler (and thus less luminous) ADs have redder colours.
Real ADs depart from the simple MTBB model because of relativistic effects and
radiation transport. However, Bonning et al. (\cite{Bonning07}) and Davis et al. (\cite{Davis07}) revealed a
similar discrepancy also for more elaborated models. Instead of a more detailed analysis, that is clearly
beyond the scope of this paper, we refer to Davis et al. for a comprehensive discussion.
These authors mention in particular that reddening by dust intrinsic to the quasar or the host galaxy and
difficulties in modelling the emission near the Balmer edge may play an important role for the discrepancies.

%-------------------------------------------------------------------------------------------------------------------
%
\subsection{The bluer-when-brighter behaviour of variability}
%
%-------------------------------------------------------------------------------------------------------------

There exists also an intrinsic slope-luminosity effect: It has
been known for a long time that quasars become bluer as they become brighter.
In Paper 1, we found that the wavelength dependence of the variability
can be expressed by $\sigma_{\rm F} \propto \lambda^{-2}$,  where $\sigma_{\rm F}$ is
the rms deviation of the flux $F_{\lambda}$ from its mean value. A similar result was reported by Wilhite et al.
(\cite{Wilhite05}) for about $300$ quasars with SDSS spectra from several epochs.
Again, this behaviour is {\it qualitatively} expected for a black-body model.

% Fig. 7
\begin{figure}[bhtp]
\includegraphics[width=6.5cm,angle=270]{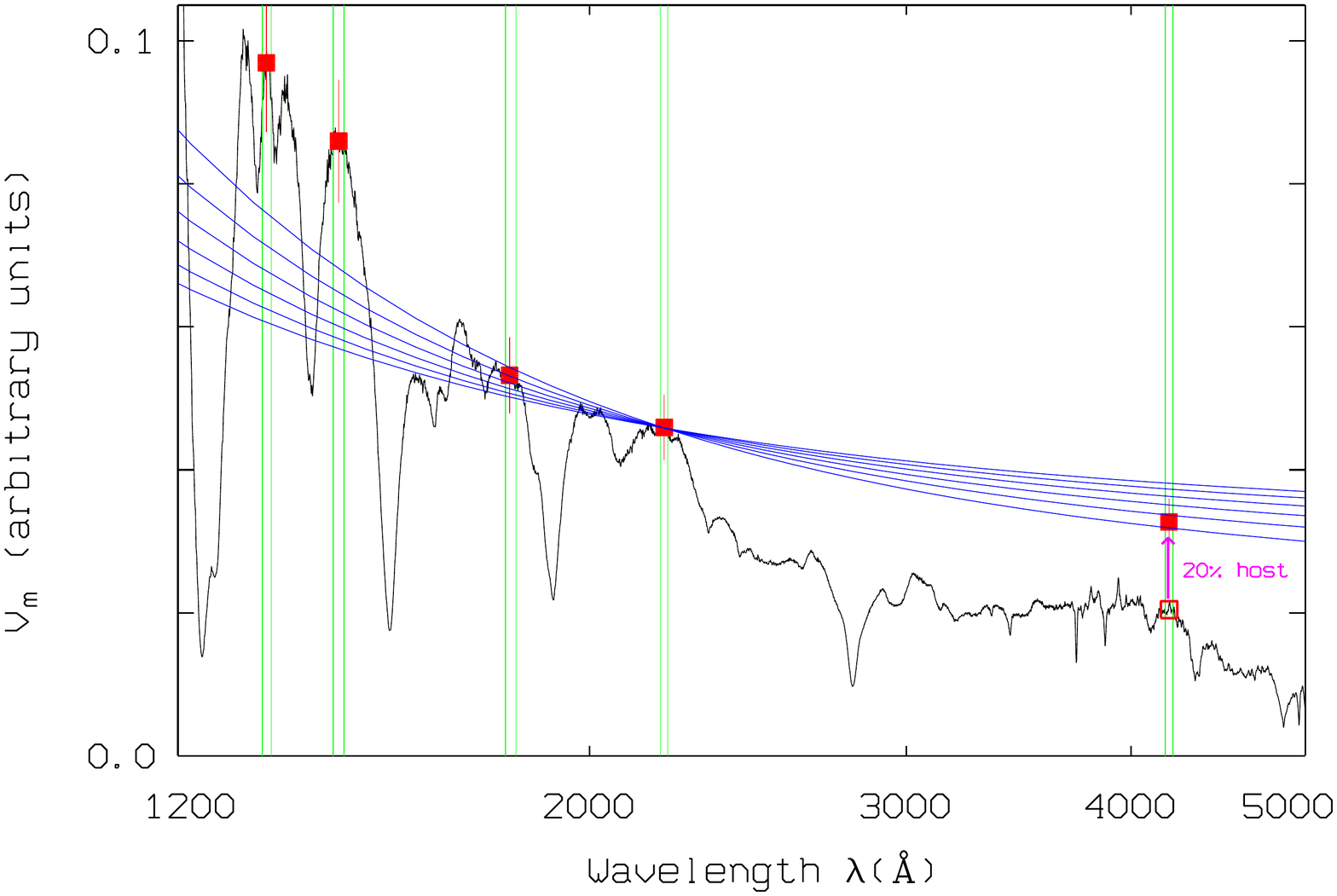}
\caption{Magnitude variance $V_m$ as a function of wavelength (black structured curve) compared with
the predictions from the MTBB models from Fig.\,\ref{fig:AD_spectra} (smooth curves). Filled squares with
error bars: mean observed $V_m$ (host corrected at $\lambda > 4000$\AA) in the continuum windows (vertical lines).}
\label{fig:var-lambda}
\end{figure}
% /qso_sdss_s82/variable_acc_rate/variance/model_10_3_0p75/plot_variance_4.prg

For the simple MTBB model the spectral slope depends
only on $T^\ast$ (Eq. \ref{eqn:T1}).
The observed spectral variability of quasars has been suggested to be explained
by sudden changes in the accretion rate, i.e., by jumps of the AD
from one stationary state to another\footnote{Or alternatively by jumps of the state of an external heating source.}
(Pereyra et al. \cite{Pereyra06}; Li \& Cao \cite{Li08}).
However, this explanation is faced with two major problems:
(i) %Variations seem to occur nearly simultaneously at different wavelengths.
Changes in the accretion flow are characterised by time-scales much longer than the time span of the
observations (Sect.\,\ref{sec:var}). Moreover, fluctuations of the
accretion rate are expected to smooth out as they move inwards.
(ii) %{\it Wavelength dependence of far UV variability}.
We simulated variable spectra by varying $\dot{M}$ in Eq.\,(\ref{eqn:T1}) and
computed the corresponding $V_m-\lambda$ relation.
Fig.\,\ref{fig:var-lambda} shows the results for the simple MTBB model ADs
from Fig.\,\ref{fig:AD_spectra}. The model curves are normalised to the observed $V_m-\lambda$
relation at $2200$\,\AA. Note that the
comparison has to be restricted to the continuum windows. After accounting for the dilution effect
from the star light of the host galaxy at $\lambda > 4000$\AA, the model predictions are roughly in
agreement with the observed relation for wavelengths longwards of the \ion{C}{iv} line.
At shorter wavelengths, however, the variability of real quasars is stronger than predicted.

%-------------------------------------------------------------------------------------------------------------------
%
\subsection{Discussion}\label{subsec:discussion}
%
%-------------------------------------------------------------------------------------------------------------
These discrepancies are difficult to explain by the simplest version of the standard AD and
seem to require an additional component. In such a two-component model, the standard AD is
the (more or less) static component, whereas a second component is responsible for the strong
variability at short wavelengths. Possible explanations include irradiation of the AD by a
central X-ray source, reprocessing of high-energy photons from the inner regions in the outer disk,
or non-intrinsic components of the observed quasar variability.

A special version of a two-component model is the strongly inhomogeneous AD model predicted
by Dexter \& Agol (\cite{Dexter11}). These authors argue that quasar variability is probably
the added effect of many, independently varying regions rather than caused by coherent
variations of the whole disk.
ADs are likely magnetised and once magnetic fields are included,
the ADs become subject to magneto-rotational instabilities that
may be the driving mechanisms behind the accretion process
(Balbus \& Hawley \cite{Balbus91}, \cite{Balbus98};
Fragile et al. \cite{Fragile07};
McKinney \& Blandford \cite{McKinney09};
Hirose et al. \cite{Hirose09}).
As was noticed already by Shakura \& Sunyaev (\cite{Shakura73}),
effects connected to magnetic fields and turbulence are
expected to be important for the radiation of the disk if the viscosity
is not too small. In this case, flares and hot spots on the surface
produce an inhomogeneous disk. The local stochastic flux variations may be a
dominant source of the observed optical/UV variability.

In their toy models, Dexter \& Agol (\cite{Dexter11}) divided the AD
into $N$ evenly spaced zones in log\,$r$ and $\varphi$. The local effective
temperature $T$ and the local radiation flux $F$
were assumed to vary with azimuth $\varphi$ at given radius $r$
and with time $t$ at given $r$ and $\varphi$. Motivated by studies
of the optical/UV variability of quasars (Kelly et al. \cite{Kelly09};
MacLeod et al. \cite{MacLeod10}), $\log\,T$ was simulated to follow
an independent damped random walk in each zone. The global,
averaged properties of the standard
thin AD model were assumed to be basically correct. The mean value of the
temperature was chosen so that the flux averaged over azimuth and time
corresponds to the standard thin disk model with
$F(r) \propto \langle T(r)\rangle^4_{t, \varphi}$.
Dexter \& Agol showed that ADs with large local temperature
fluctuations $\sigma_T \approx 0.4$\,dex in $N \approx 10^2-10^3$ independently
varying zones can explain the observed discrepancy of the AD size (Sect.\,1)
while matching the observed random walk characteristics of the optical/UV
variability.

These local instabilities affect the accretion flow on the one hand and the
radiation flux variations on the other. We thus naturally expect a relation between
the UV variability and the accretion rate. Dexter \& Agol (\cite{Dexter11}) modelled the time-scale 
and the amplitudes of the temperature fluctuations as independent of $r$. If we assume that these
characteristics are also independent of the BH mass (and spin),
the quasar-to-quasar differences in the observed variability
may be related to different numbers $N$ of
independently varying zones. As the variance decreases with $N$, we have to expect that
more variable quasars have ADs with smaller $N$ and that the variability is consequently
anti-correlated with $\dot{M}$.

%%%%%%%%%%%%%%%%%%%%%%%%%%%%%%%%%%%%%%%%%%%%%%%%%%%%%%%%%%%%%%%%%%%%%%%%%%%%%%%%%%%%%%%%%%%%%%%%%%%%%%%%%%
%
%
\section{Correlations between variability, accretion rate, and black hole mass}\label{sec:correlations}
%
%
%%%%%%%%%%%%%%%%%%%%%%%%%%%%%%%%%%%%%%%%%%%%%%%%%%%%%%%%%%%%%%%%%%%%%%%%%%%%%%%%%%%%%%%%%%%%%%%%%%%%%%%%%%

% Fig. 8 a-t
\begin{figure*}[bhtp]
\begin{tabbing}
\includegraphics[viewport=12 0 440 610,scale=0.265,clip]{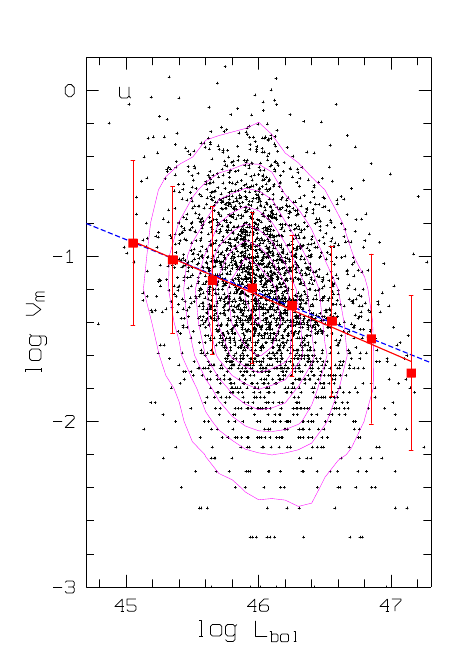}\hfill \=
\includegraphics[viewport=70 0 440 610,scale=0.265,clip]{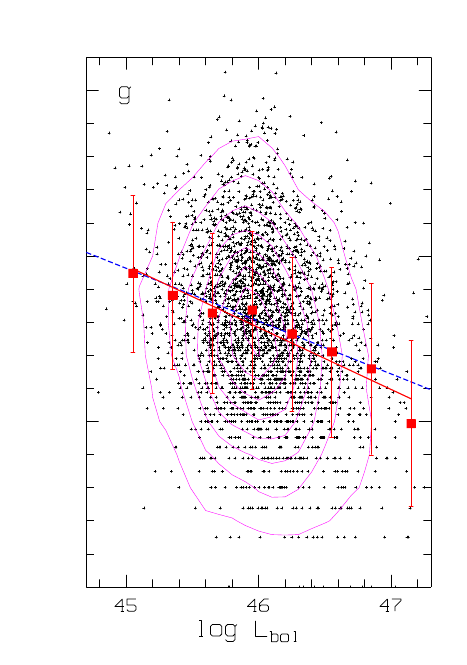}\hfill \=
\includegraphics[viewport=70 0 440 610,scale=0.265,clip]{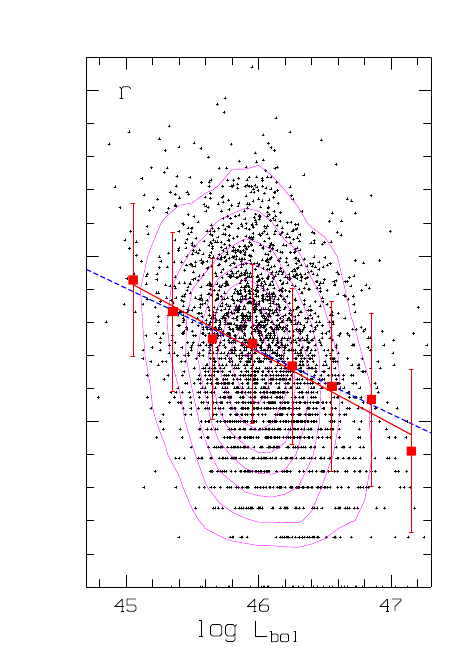}\hfill \=
\includegraphics[viewport=70 0 440 610,scale=0.265,clip]{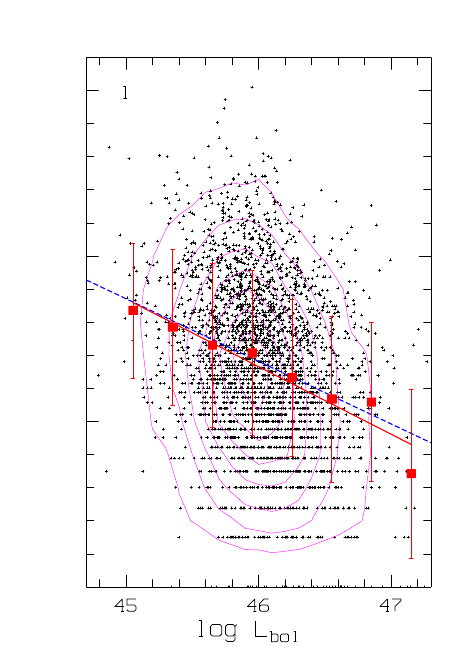}\hfill \=
\includegraphics[viewport=70 0 440 610,scale=0.265,clip]{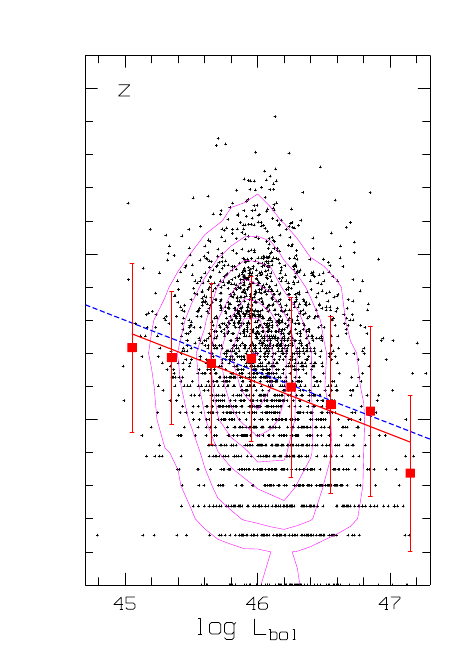}\hfill \\
\includegraphics[viewport=12 0 440 610,scale=0.265,clip]{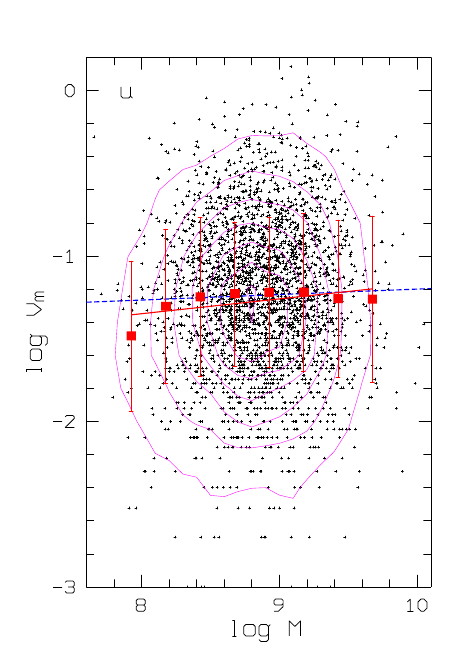}\hfill \=
\includegraphics[viewport=70 0 440 610,scale=0.265,clip]{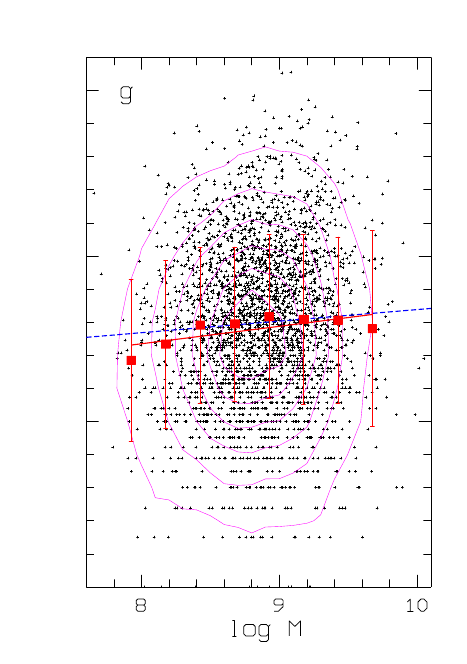}\hfill \=
\includegraphics[viewport=70 0 440 610,scale=0.265,clip]{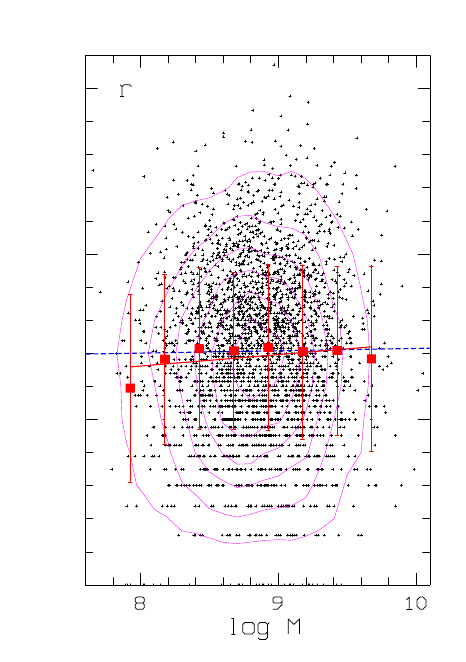}\hfill \=
\includegraphics[viewport=70 0 440 610,scale=0.265,clip]{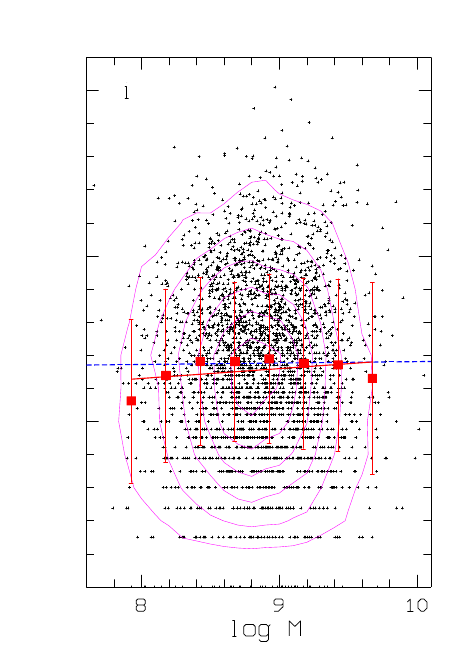}\hfill \=
\includegraphics[viewport=70 0 440 610,scale=0.265,clip]{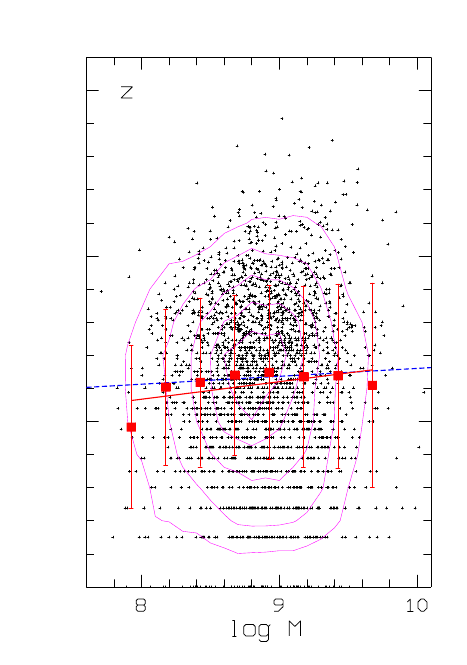}\hfill \\
\includegraphics[viewport=12 0 440 610,scale=0.265,clip]{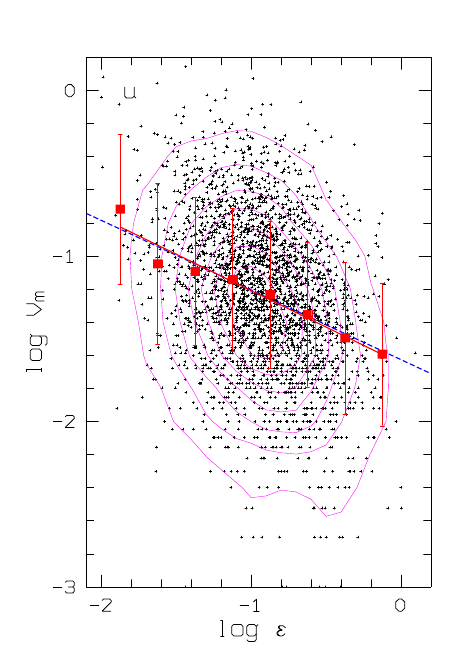}\hfill \=
\includegraphics[viewport=70 0 440 610,scale=0.265,clip]{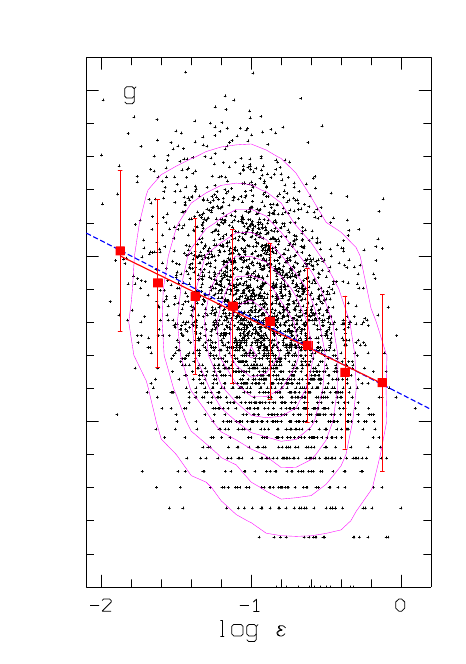}\hfill \=
\includegraphics[viewport=70 0 440 610,scale=0.265,clip]{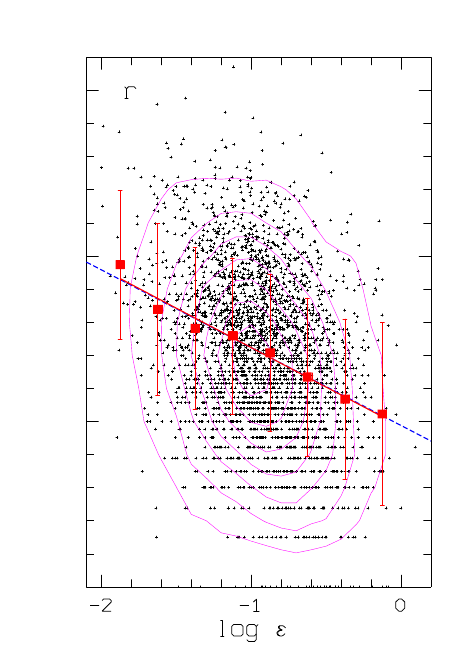}\hfill \=
\includegraphics[viewport=70 0 440 610,scale=0.265,clip]{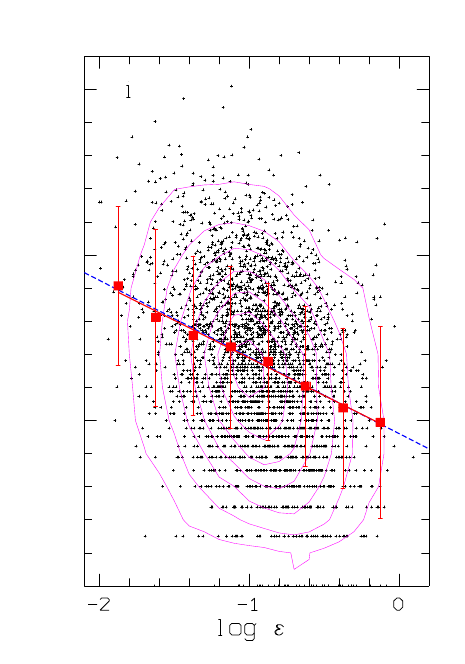}\hfill \=
\includegraphics[viewport=70 0 440 610,scale=0.265,clip]{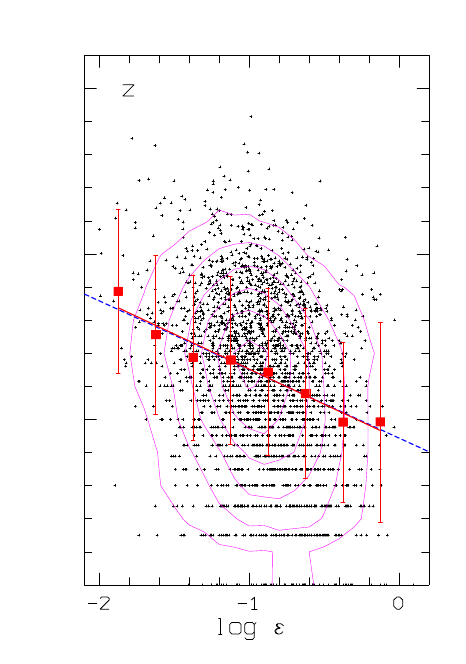}\hfill \\
\includegraphics[viewport=12 0 440 610,scale=0.265,clip]{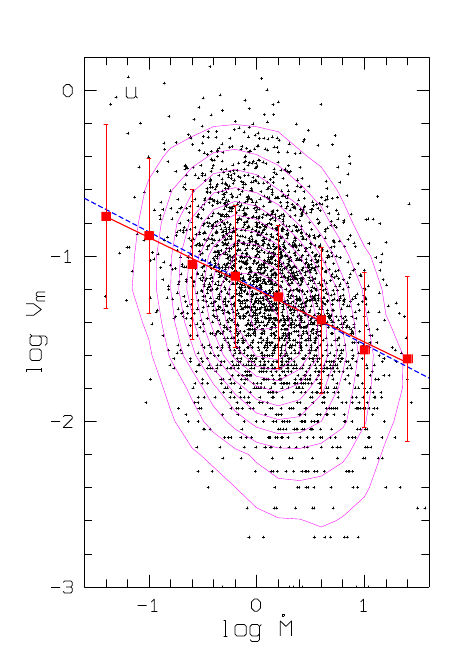}\hfill \=
\includegraphics[viewport=70 0 440 610,scale=0.265,clip]{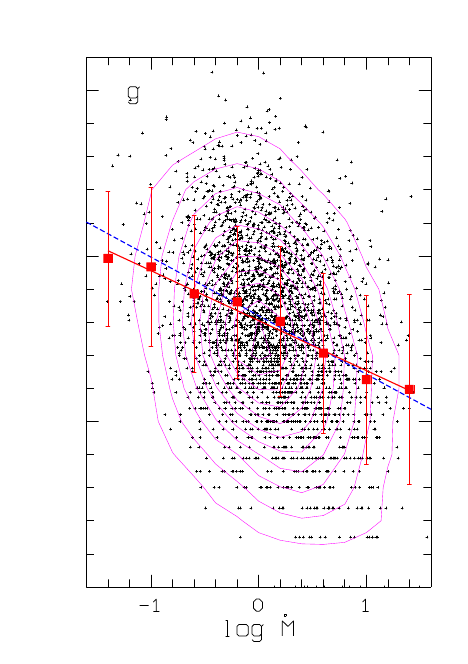}\hfill \=
\includegraphics[viewport=70 0 440 610,scale=0.265,clip]{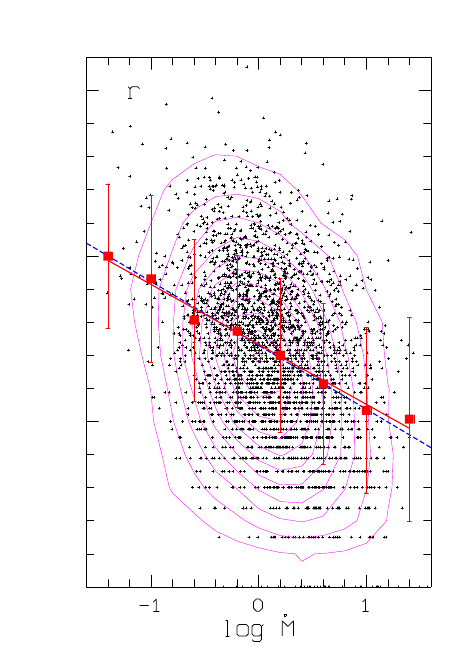}\hfill \=
\includegraphics[viewport=70 0 440 610,scale=0.265,clip]{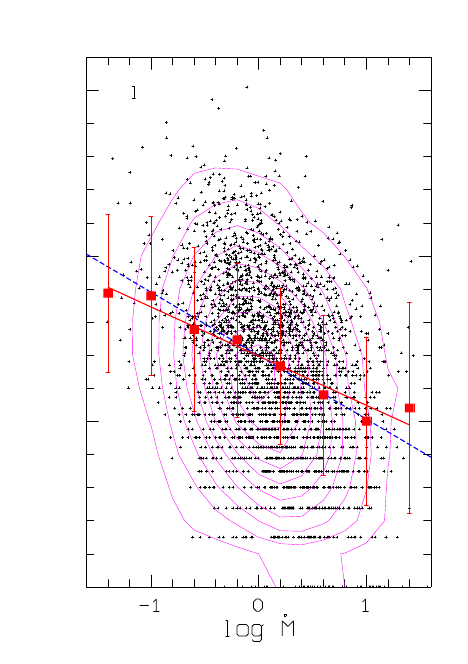}\hfill \=
\includegraphics[viewport=70 0 440 610,scale=0.265,clip]{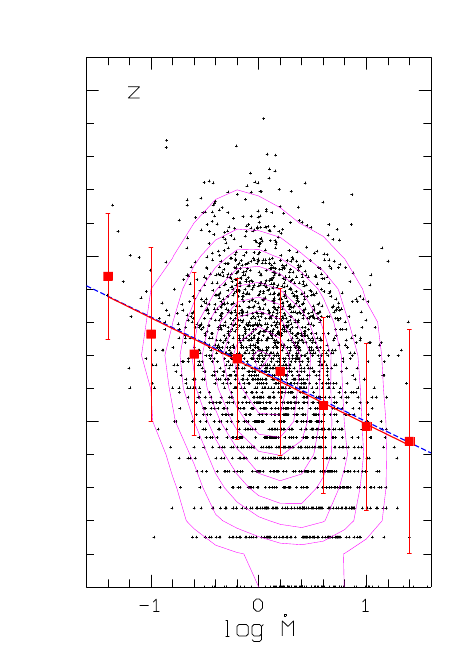}\hfill
\end{tabbing}
\caption{
Variability estimator $V_m$ versus $L_{\rm bol}, M, \varepsilon$,
and $\log\,\dot{M}$ for the five SDSS bands. Over-plotted are  density contours
and the linear regression lines for binned (solid) and unbinned (dashed) data.
}
\label{fig:regr_all}
\end{figure*}
%Weiss/Shen+11/regressions/plot_regressions_all_with_contours.prg

% Fig. 9 a-t
\begin{figure*}[bhtp]
\begin{tabbing}
\includegraphics[viewport=10 0 300 430,scale=0.382,clip]{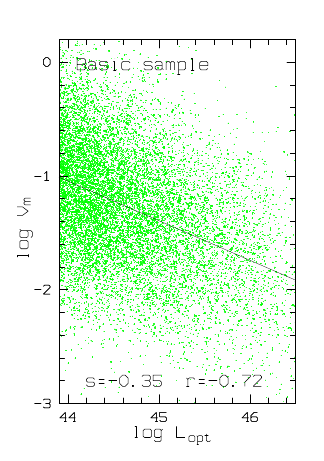}\hfill \=
\includegraphics[viewport=45 0 300 430,scale=0.382,clip]{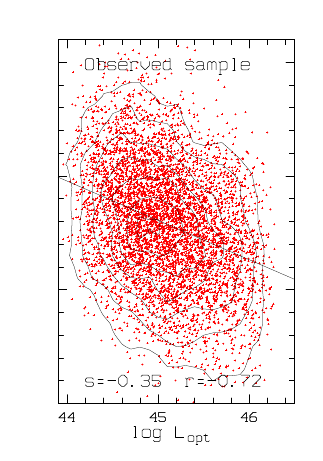}\hfill \=
\includegraphics[viewport=45 0 300 430,scale=0.382,clip]{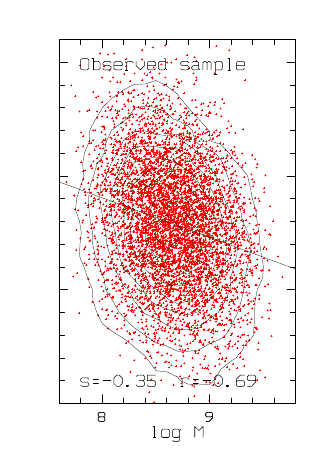}\hfill \=
\includegraphics[viewport=45 0 300 430,scale=0.382,clip]{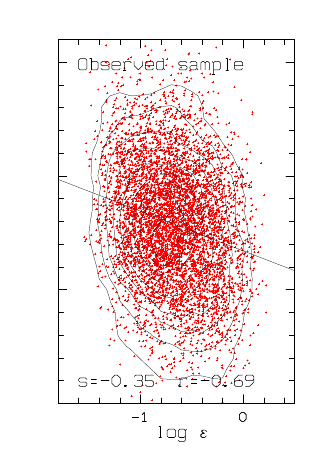}\hfill \=
\includegraphics[viewport=45 0 300 430,scale=0.382,clip]{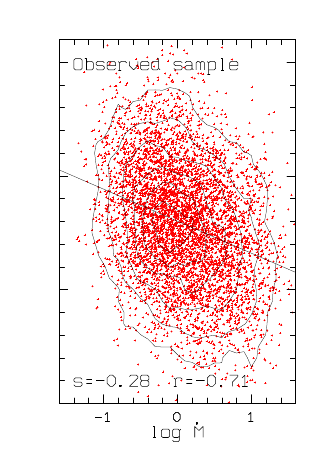}\hfill \\
\includegraphics[viewport=10 0 300 430,scale=0.382,clip]{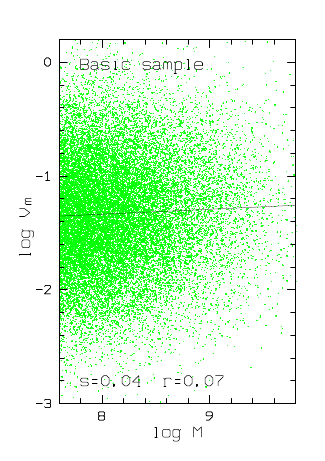}\hfill \=
\includegraphics[viewport=45 0 300 430,scale=0.382,clip]{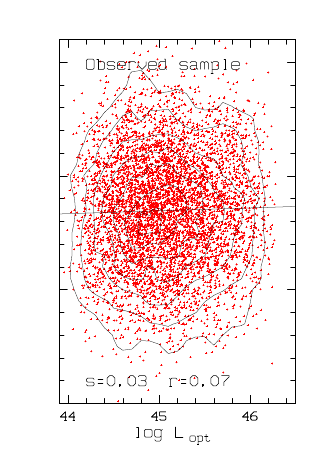}\hfill \=
\includegraphics[viewport=45 0 300 430,scale=0.382,clip]{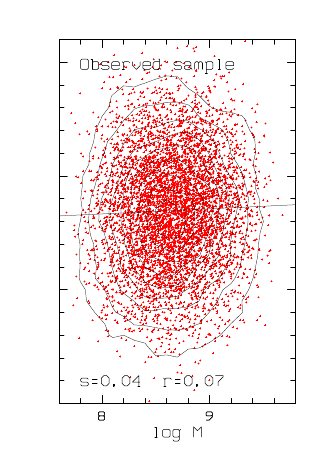}\hfill \=
\includegraphics[viewport=45 0 300 430,scale=0.382,clip]{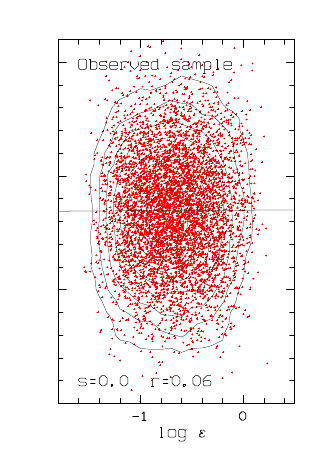}\hfill \=
\includegraphics[viewport=45 0 300 430,scale=0.382,clip]{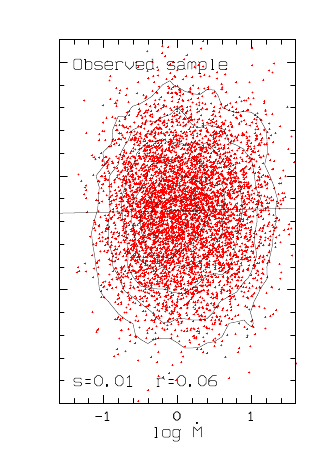}\hfill \\
\includegraphics[viewport=10 0 300 430,scale=0.382,clip]{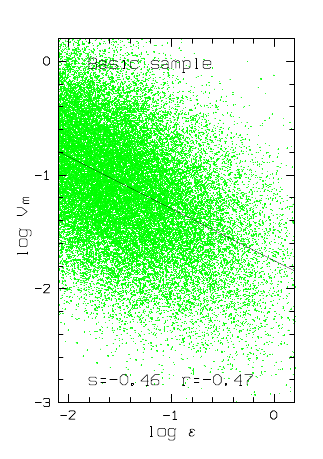}\hfill \=
\includegraphics[viewport=45 0 300 430,scale=0.382,clip]{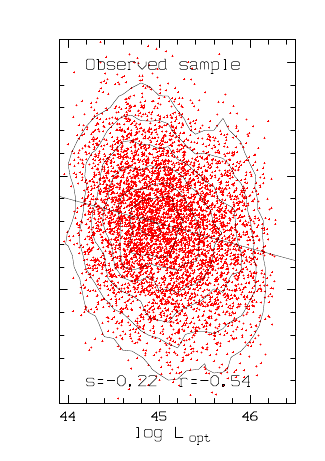}\hfill \=
\includegraphics[viewport=45 0 300 430,scale=0.382,clip]{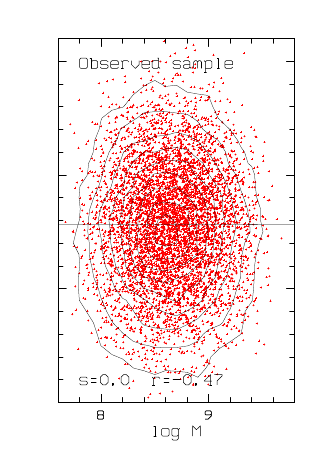}\hfill \=
\includegraphics[viewport=45 0 300 430,scale=0.382,clip]{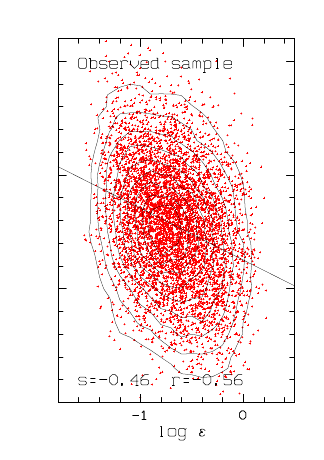}\hfill \=
\includegraphics[viewport=45 0 300 430,scale=0.382,clip]{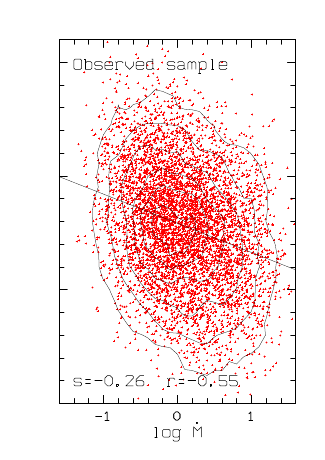}\hfill \\
\includegraphics[viewport=10 0 300 430,scale=0.382,clip]{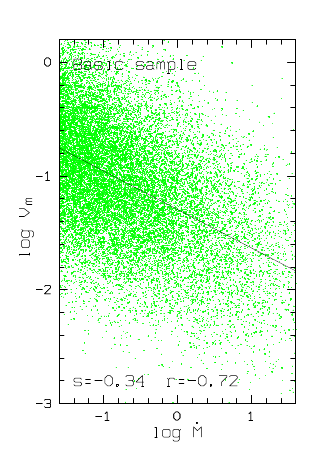}\hfill \=
\includegraphics[viewport=45 0 300 430,scale=0.382,clip]{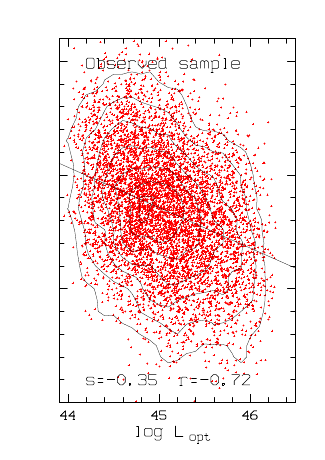}\hfill \=
\includegraphics[viewport=45 0 300 430,scale=0.382,clip]{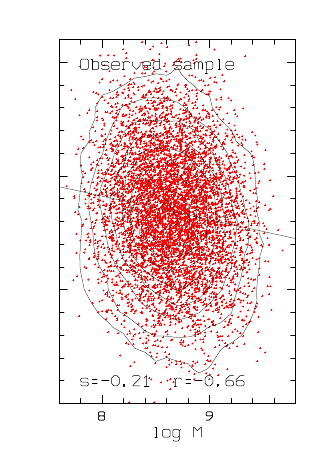}\hfill \=
\includegraphics[viewport=45 0 300 430,scale=0.382,clip]{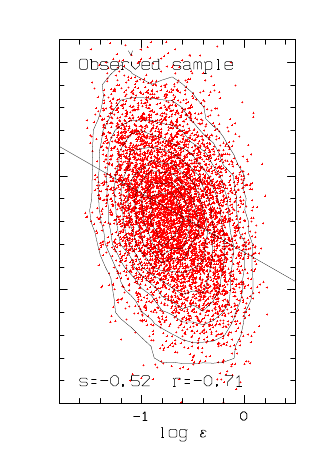}\hfill \=
\includegraphics[viewport=45 0 300 430,scale=0.382,clip]{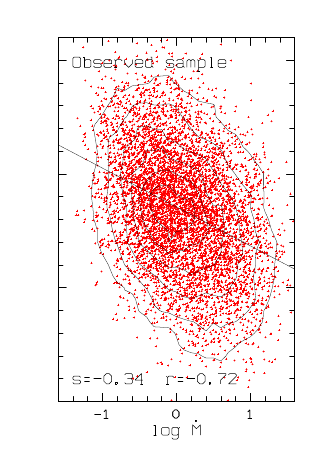}\hfill
\end{tabbing}
\caption{
Variability estimator $V_m$ versus $L_{\rm opt}, M, \varepsilon,$ and $\dot{M}$ for the
simulated quasar sample. In each row, the left panel shows the input correlation in the basic sample, the other
diagrams show the expected correlations in the ``observed'' simulated sample with density contours over-plotted.
The two numbers close to the bottom
of each panel are the slope (s) and the regression coefficient (r) from the linear regression (solid).
}
\label{fig:regr_all_sim}
\end{figure*}
%Weiss/Shen+11/simulations/variability/plot_var_sim_all.prg

%-------------------------------------------------------------------------------------------------------------------
%
\subsection{Observed correlations}\label{subsec:corr_obs}
%
%-------------------------------------------------------------------------------------------------------------

To search for possible correlations between the variability and the accretion parameters, we restricted
the sample to those quasars with small measurement uncertainties of
$< 0.25$\,dex in the fiducial virial BH mass.
As pointed out by Shen et al. (\cite{Shen11}; see also Sect.\,\ref{sec:BH_mass}), the mass uncertainty was
propagated from the measurement uncertainties of the continuum luminosities and the line widths, but does neither
include systematic effects nor the statistical uncertainty from the calibration of the scaling relations
which is probably larger than the 0.25\,dex.
We also removed a small number (66) of broad absorption line (BAL) quasars (BAL flag $> 0$) from the sample
because BALs can strongly affect the mass estimation. Finally, 68 radio-loud quasars 
($R \equiv F_{\rm 6cm}/F_{\rm 2500\AA} > 10$)
were removed because their variability may be influenced, or even dominated, by the non-thermal component.
These restrictions result in a final sample of 2951 quasars. It should be noticed, however, that these additional
limitations have an only marginal effect on the results.

\begin{table}[h]
\caption{Slopes and test statistics for the correlations between $\mbox{log}\,V_m$ and
four other quantities from Fig.\,\ref{fig:regr_all}.
}
\centering
\begin{tabular}{lcccccc}
\hline\hline
%&&\\
Quantity           &Band&Slope& $r$      &  $t$  &$D_{\rm max}$ & $D_{\rm crit, \alpha=0.99}$ \\
(1)                &(2) & (3) & (4)      &  (5)  & (6)          &  (7) \\
%&&&\\
\hline
%&&&\\
$\log L_{\rm bol} $&  u &-0.31&  $-0.24$ &  13.2 & 0.242 & 0.074 \\
$\log L_{\rm bol} $&  g &-0.30&  $-0.23$ &  12.5 & 0.196 & 0.073 \\
$\log L_{\rm bol} $&  r &-0.37&  $-0.27$ &  15.0 & 0.240 & 0.073 \\
$\log L_{\rm bol} $&  i &-0.37&  $-0.27$ &  14.9 & 0.240 & 0.073 \\
$\log L_{\rm bol} $&  z &-0.30&  $-0.19$ &  10.7 & 0.216 & 0.081 \\
 \hline
$\log M  $&  u &+0.05&  $ 0.04$ &  2.03 & 0.047 & 0.074 \\
$\log M  $&  g &+0.09&  $ 0.06$ &  3.43 & 0.061 & 0.073 \\
$\log M  $&  r &+0.03&  $ 0.02$ &  1.07 & 0.025 & 0.073 \\
$\log M  $&  i &+0.02&  $ 0.01$ &  0.78 & 0.028 & 0.073 \\
$\log M  $&  z &+0.05&  $ 0.04$ &  1.98 & 0.060 & 0.080 \\
 \hline
$\log \varepsilon $&  u &-0.41&  $-0.30$ &  16.8 & 0.268 & 0.074 \\
$\log \varepsilon $&  g &-0.46&  $-0.31$ &  17.7 & 0.258 & 0.073 \\
$\log \varepsilon $&  r &-0.47&  $-0.31$ &  17.7 & 0.272 & 0.073 \\
$\log \varepsilon $&  i &-0.46&  $-0.30$ &  17.2 & 0.252 & 0.073 \\
$\log \varepsilon $&  z &-0.41&  $-0.25$ &  13.9 & 0.235 & 0.080 \\
 \hline
$\log \dot{M} $    &  u &-0.33&  $-0.32$ &  18.3 & 0.474 & 0.074 \\
$\log \dot{M} $    &  g &-0.34&  $-0.32$ &  18.2 & 0.610 & 0.073 \\
$\log \dot{M} $    &  r &-0.38&  $-0.34$ &  19.8 & 0.254 & 0.073 \\
$\log \dot{M} $    &  i &-0.38&  $-0.34$ &  19.4 & 0.281 & 0.073 \\
$\log \dot{M} $    &  z &-0.31&  $-0.25$ &  13.9 & 0.603 & 0.080 \\
%&&&\\
\hline
\end{tabular}
\label{tab:corr}
\tablefoot{ Columns 4 and 5: correlation coefficient $r$ and test statistic $t$,
columns 6 and 7: KS test statistics $D_{\rm max}$ and its critical value $D_{\rm crit}$.}
\end{table}
%Weiss/Shen+11/regressions/regression_all_z.tbl

Figure\,\ref{fig:regr_all} shows the double-logarithmic diagrams of the variability estimator $V_m$ in the five SDSS bands ugriz
(left to right) versus (top to bottom) the bolometric luminosity, the BH mass, the Eddington ratio, and the accretion rate. Over-plotted
are equally spaced density contours and regression curves. Linear regressions were performed for the unbinned data, but also
for the mean values in 8 binning intervals. The two regression curves are generally in good agreement with each other.
The Pearson product-moment correlation coefficients $r$
for the unbinned date are listed in Table\,\ref{tab:corr} along with the test statistics $t = r \sqrt{(N-2)/(1-r^2)}$.
The null hypothesis $H_0$, that there is no correlation, has to be rejected on a significance level of 99\% if
$t>t_{\rm crit, \alpha=0.99} = 2.58$. Though the scatter is large in all panels of Fig.\,\ref{fig:regr_all}, it is evident
from Table\,\ref{tab:corr} that $H_0$ can be rejected both for $\log L_{\rm bol}$, $\log \varepsilon$, and $\log \dot{M}$
in all five SDSS bands. Highest confidence is indicated for
$\log \dot{M}$ followed by $\log \varepsilon$. Because
$\dot{M}$ and $\varepsilon$ are strongly correlated with each other (Fig.\,\ref{fig:eta_M}),
we conclude that our analysis reveals a significant anti-correlation between
variability and accretion rate. The correlation between variability and BH mass is marginally significant in 
the g band only, whereas the null hypothesis has not to be rejected for the other bands.

To consolidate the results from the correlation test, we applied
yet another statistical test: We divided the range of  $\log \dot{M}$ into three intervals with about the same
number of quasars per interval defining thus three sub-samples S$_1$, S$_2$, and S$_3$ with
$\log \dot{M}(\mbox{S}_1) < \log \dot{M}(\mbox{S}_2) < \log \dot{M}(\mbox{S}_3)$. Now, we compared the distributions of
$\log V_m$ in the sub-samples S$_1$ and S$_3$ with each other. The Kolmogorov-Smirnov test was used to check
the null hypothesis $H_0$ that both distributions are statistically indistinguishable. $H_0$ has to be rejected
at level $\alpha$ if the test statistic $D_{\rm max}$ is greater than a critical value $D_{\rm crit}(\alpha,n_1,n_3)$,
where $n_1$ and $n_3$ are the numbers of quasars in S$_1$ and S$_3$, respectively. The same test was applied also
to $\log L_{\rm bol}, \log M$, and $\log \varepsilon$. The test statistic $D_{\rm max}$ and the critical values
$D_{\rm crit}$ for $\alpha=0.99$ are listed in Table\,\ref{tab:corr}. Obviously, we have to reject the hypothesis that
the variabilities of the low-$\dot{M}$ quasars and the high-$\dot{M}$
quasars come from the same distribution. The same conclusion can be drawn for $\varepsilon$ and $L_{\rm bol}$, but not for
$M$. Highest confidence is found for $\dot{M}$. The results from the correlation test are therewith
clearly confirmed. Finally, we applied the Mann-Whitney $U$ test to the same sub-samples. Though
less sensitive to the presence of outliers, this test does not alter the conclusion derived from the Kolmogorov-Smirnov test.

\begin{figure}[bhtp]
\begin{tabbing}
\includegraphics[viewport=1 95 550 790,scale=0.24,clip]{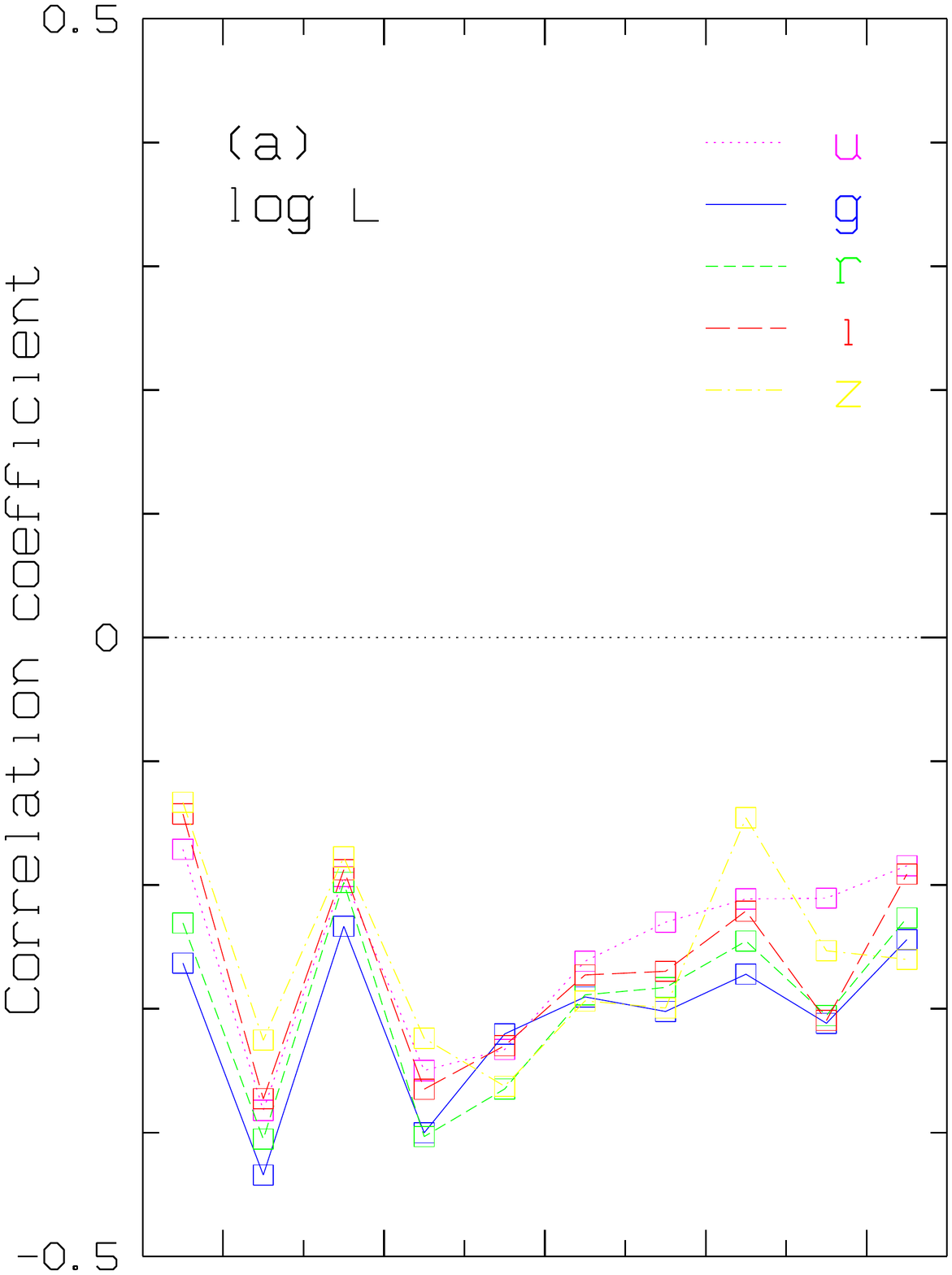}\hfill \=
\includegraphics[viewport=90 95 550 790,scale=0.24,clip]{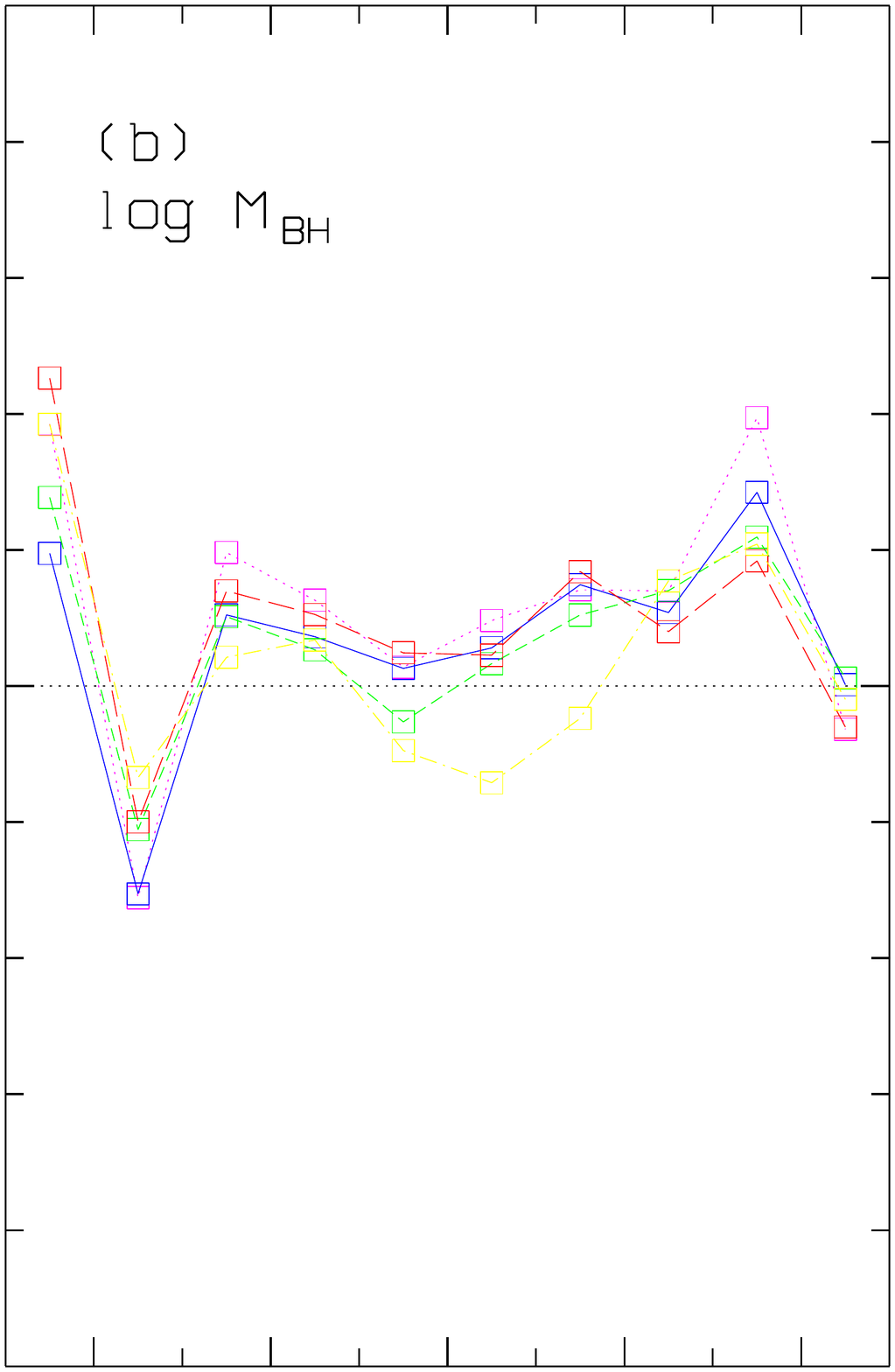}\hfill \\
\includegraphics[viewport=1 10 550 780,scale=0.24,clip]{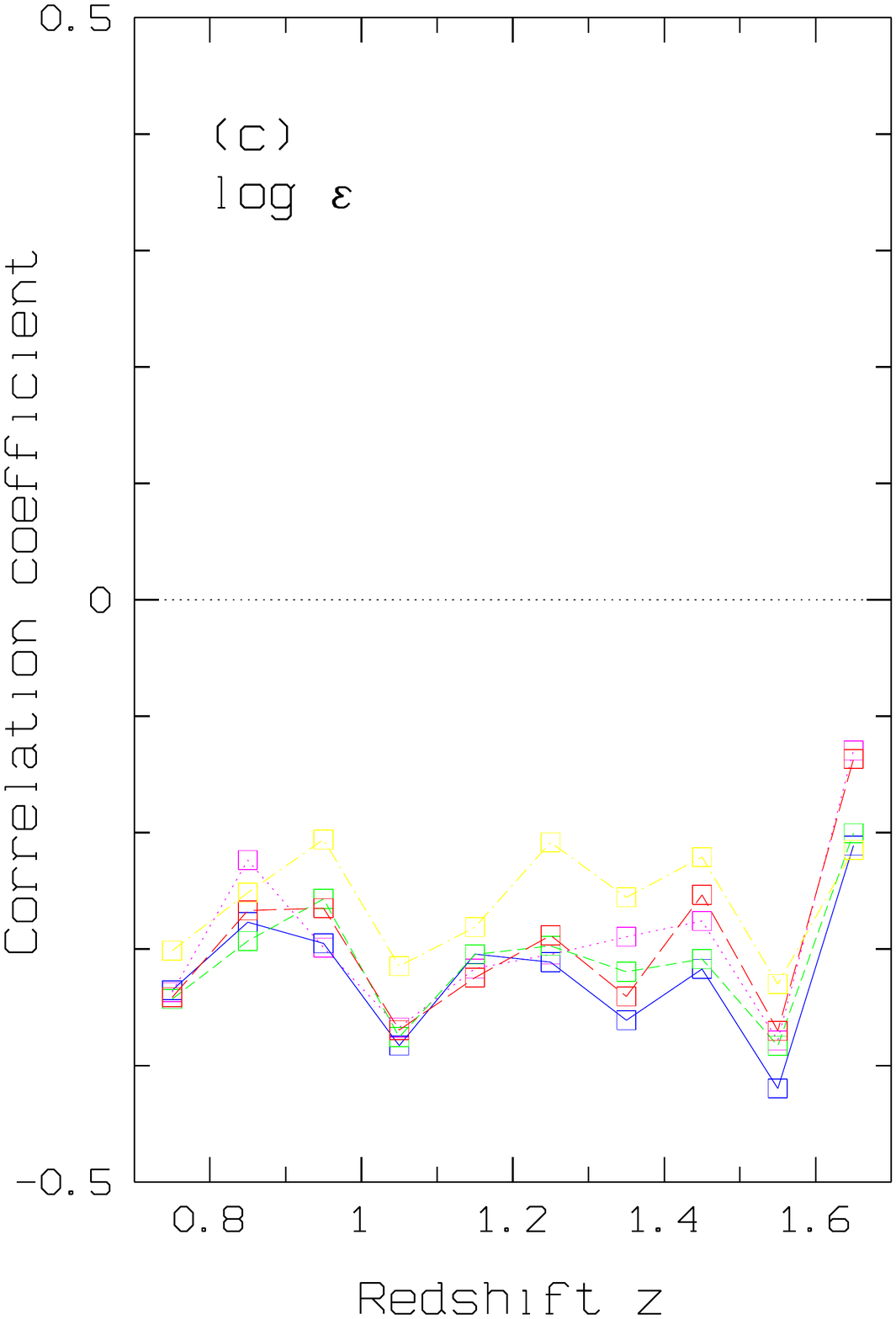}\hfill \=
\includegraphics[viewport=90 10 550 780,scale=0.24,clip]{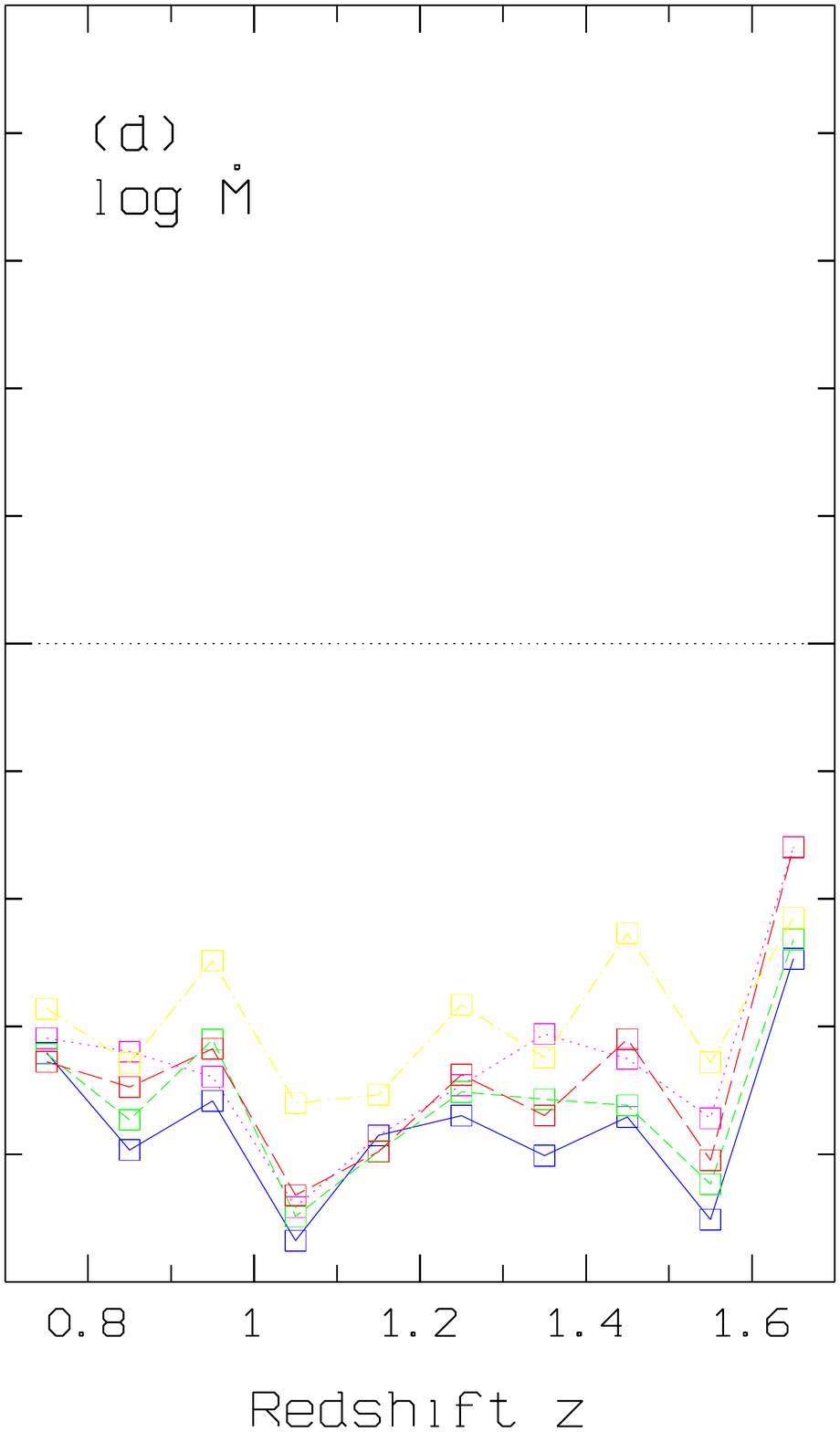}\hfill
\end{tabbing}
\caption{
Correlation coefficients for the relationships between
the variability $\log\,V_m$ in the five SDSS bands and
$\log\,L_{\rm bol}$ (a), $\log\,M$ (b), $\log\,\varepsilon$ (c), and $\log\,\dot{M}$ (d)
in $z$-binned data.
}
\label{fig:corr_coeffs}
\end{figure}

We checked the impact of various factors that may potentially influence the correlations. First,
we computed the correlation coefficients for the relations from Fig.\,\ref{fig:regr_all} in $z$ intervals
of the width $\Delta z = 0.1$. Figure\,\ref{fig:corr_coeffs} shows that there is no significant
redshift dependence of $r$. Interestingly $r$ is smaller than average (Table\,\ref{tab:corr}) for
$\log L_{\rm bol}$, $\log \varepsilon$, and $\log \dot{M}$ in the bins at $z=1.05$ and 1.55 where the 
$z$ distribution has local maxima (Fig.\,\ref{fig:mass_error})
reaching $r \approx -0.5$ for $\log \dot{M}$. Secondly,
the influence of outliers
was checked by a sigma clipping and was found to be negligible. Thirdly, the results are found to be
essentially independent against the inclusion of radio-loud quasars and BAL quasars. Fourthly and lastly,
we removed the quasars with low significance of the measured variability (e.g. $\chi^2 < 3$; see Paper 1)
and found again only a negligible effect on the correlations.

% Fig. 10 a-d
%Weiss/Shen+11/regressions/plot_regressions_in_bins_korrkoeff.prg

%-------------------------------------------------------------------------------------------------------------------
%
\subsection{Discussion}\label{subsec:corr_disc}
%
%-------------------------------------------------------------------------------------------------------------

The findings of statistically significant correlations does not guarantee that they are really intrinsic.
In Sect.\,\ref{sec:acc_rate}, we demonstrated via simulations that significant correlations can 
be found in a selected quasar sample even though the same properties are uncorrelated in the 
underlying parent sample, i.e. the observed correlation can be the result of selection effects in 
combination with measurement errors. 
Here we applied similar simulations to evaluate the role of selection effects on
the correlations found for the variability. We assigned a value of $\log V_{\rm m}$ to each quasar
in the basic sample where $\log V_{\rm m}$ is assumed to be Gaussian distributed with a standard deviation
$\sigma_{\log M} = 0.5$ around the mean value  $\overline{\log V_{\rm m}}$.
We started with a model where $\overline{\log V_{\rm m}}$ is uncorrelated with any of the 
studied properties in the basic sample and found no correlations for $\log V_{\rm m}$ in the 
``observed'' sample. In the next step, we simulated four different basic quasar samples for the 
assumption that $\log V_{\rm m}$ is intrinsically correlated with either 
$\log L_{\rm bol}, \log M, \log \varepsilon,$ or $\log \dot{M}$ adopting the slopes found for
the SDSS sample (Tab.\ref{tab:corr} and Fig.\,\ref{fig:regr_all}). The results are
shown in Fig.\,\ref{fig:regr_all_sim} where the four rows correspond to the four different
samples. In each row, the panel on the left-hand side shows the adopted intrinsic correlation 
in the basic sample, the other diagrams show the resulting relations in the ``observed''
(simulated) sample with the linear regression curves and the values for the slope (s) and the
regression coefficient (r) close to the bottom of each panel. 
The adopted correlation with $\log M$ in the second row is very weak
and this case is thus close to the case of uncorrelated variability mentioned above.  
For the other cases, however, the ``observed'' samples show similar distributions as 
the SDSS quasar sample and their correlations are similar to those in their basic samples. 
The best agreement is found for the intrinsic relations of the variability with the 
Eddington ratio and with the accretion rate, respectively. 
We conclude that the observed relations in Fig.\,\ref{fig:regr_all} are not produced or significantly 
influenced by the selection effects. 

A more serious problem is an obvious inconsistency in the present study. On the one hand, 
the standard AD model was used to derive the accretion rates. On the other hand, however,
we have argued that several observational facts are not well reproduced by the standard AD model.
As discussed in Sect.\,\ref{sec:slope-luminosity}, most of these discrepancies seem to 
require an additional component that may affect particularly the observed properties  
from the FUV to X-rays. The approach suggested by DL11 and adopted in the present study to estimate
$\dot{M}$ is based on the optical luminosity, i.e. on a spectral region where the observed flux is 
less influenced by regions emitting the highest energy photons. Similarly,
the discrepancy between the observed and the expected variability amplitudes is found 
at $\lambda \la 1500$\AA\ whereas the average redshift $\bar{z} = 1.24$ of our quasars guarantees that
the measured variabilities refer to longer wavelengths, particularly for the high-throughput 
SDSS bands g, r, and i. In other words, the variability data are mainly restricted to the
intrinsic wavelength range where the observed $V-\lambda$ relation can be explained by the standard model.
There remains, however, the problem that the observed time-scales seem to disfavour jumps of the AD from one
stationary state to another as the major source of variability (Sect.\,\ref{sec:slope-luminosity}).
In the standard model, the relationship between $V_m$ and $\dot{M}$ depends
on the fluctuation spectrum of $\dot{M}$ and on the wavelength. For simplicity, 
we simulated ADs where the variations $\delta \dot{M}$ are scaled with the mean accretion rate $\dot{M}$,
i.e. $|\delta \dot{M}| = 0...\Delta \dot{M}_{\rm max}$ where $\Delta \dot{M}_{\rm max} \propto \dot{M}$.
We found $\log V_m \propto c\,\log \dot{M}$ with $ c \approx -0.15...-0.07$ for $\lambda = 2000-4000$\AA\
whereas the observed variability shows much steeper relations with $c = -0.38...-0.34$ in the three 
high-throughput SDSS bands (Table\,\ref{tab:corr}). 

Following e.g. Dexter \& Agol (\cite{Dexter11}), we argued here that the global
properties of the standard thin AD model may be basically correct so that $\dot{M}$
from Eq.\,(\ref{eqn:AR_DL}) can be used at least as a proxy for the real accretion rate.
The strongly inhomogeneous AD model predicted by Dexter \& Agol 
invokes a stochastic component that may explain both the observed stochastic nature and the 
time-scales of quasar variability. We argued (Sect.\,\ref{subsec:discussion}) that the variability
should be anti-correlated with the accretion rate for such a model, qualitatively in agreement with the
results presented here.

%%%%%%%%%%%%%%%%%%%%%%%%%%%%%%%%%%%%%%%%%%%%%%%%%%%%%%%%%%%%%%%%%%%%%%%%%%%%%%%%%%%%%%%%%%%%%%%%%%%%%%%%%%
%
%
\section{Summary and conclusions}\label{sec:conclusions}
%
%
%%%%%%%%%%%%%%%%%%%%%%%%%%%%%%%%%%%%%%%%%%%%%%%%%%%%%%%%%%%%%%%%%%%%%%%%%%%%%%%%%%%%%%%%%%%%%%%%%%%%%%%%%%

The main aims of this study were to estimate the accretion rates for a suitable sample of quasars and to
search for correlations of the UV quasar variability with the fundamental parameters of the accretion process.
We have compiled a catalogue of about $4000$ quasars including individual variability estimators from Paper 1 
(derived from the
multi-epoch photometry in the SDSS Stripe 82) and both luminosities and virial black hole mass estimates $M$
from Shen et al. (\cite{Shen11}). The selection was restricted to the redshift interval $0.7 \le z \le 1.7$
where the black hole masses are estimated from the \ion{Mg}{ii} line and the mass errors are believed to be
smallest (Fig.\,\ref{fig:mass_error}).
We computed the mass accretion rates $\dot{M}$ from the optical continuum luminosity 
$L_{\rm opt}$ and the black hole mass $M$, where $L_{\rm opt}$ was extrapolated from 
$L_{\rm 3000}$ adopting a power-law continuum with individually
measured spectral indexes $\alpha_\lambda$. The analysis leads to the following conclusions:

\begin{itemize} 
\item[1.]
The quasar composite spectra from FUV to NIR are in accordance with the predictions from the standard 
AD model (Fig.\,\ref{fig:AD_spectra}). On the other hand, the individually determined spectral
slopes $\alpha_\lambda$ do not show the expected dependence on $T^\ast$ (Fig.\,\ref{fig:alpha_T}). 
This result qualitatively confirms earlier findings by Bonning et al. (\cite{Bonning07})
and by Davis et al. (\cite{Davis07}).
\item[2.]
The wavelength dependence of the strength of the variability, $V(\lambda)$, is in agreement with the predictions 
from the standard thin AD model for wavelengths longwards of the \ion{C}{iv} line. 
At shorter wavelengths, the observed variability is stronger than predicted by the model
(Fig.\,\ref{fig:var-lambda}). In the standard model, variability is explained by sudden changes
of the accretion rate, but it is well known that such an interpretation is faced with a time-scale 
problem (e.g. MacLeod et al. \cite{MacLeod10}).
\item[3.]
Despite the problems with the standard AD, we argued here that the global properties of the
model may be basically correct and that the model predictions can be used to estimate $\dot{M}$.
We applied the scaling relation from DL11 to derive $\dot{M}$ from the optical luminosity and $M$.
We found that $\dot{M}$ is correlated with the Eddington ratio
$\varepsilon$,  $L_{\rm bol}$, but not significantly with $M$ (Fig.\,\ref{fig:eta_M}).
We estimated the radiative efficiency $\eta$ assuming a constant bolometric correction and
found an $\eta-M$ relation (Fig.\,\ref{fig:eta_M}) very similar to that one reported by DL11 
(see also Laor \& Davis \cite{Laor11a}) from their study of a smaller sample of lower-$z$ 
Palomar-Green quasars. We also found  $\eta$ to be tightly anti-correlated with $\varepsilon$.
We studied these relations via simple numerical simulations and concluded that they can be
explained as artificial caused mainly by the selection effects in combination with the mass
errors. A broadly similar result was recently reported by Wu et al. (\cite{Wu13}).
\item[4.]
We confirmed the existence of significant anti-correlations of the variability estimator $V_m$ in the five
SDSS bands with both $\varepsilon$ and $L_{\rm bol}$, but not with $M$ (Fig.\,\ref{fig:regr_all};
Table\,\ref{tab:corr}). Our study reveals also, for the first time, an anti-correlation of $V_m$
and $\dot{M}$.
Based on numerical simulations we argued that these anti-correlations are not produced by the selection
effects. On the other hand, $L, \varepsilon$, and $\dot{M}$ correlate with each other and it is thus
difficult to decide which of these quantities is intrinsically correlated with $V_{\rm m}$ and which
correlations of $V_{\rm m}$ are produced by the $L - \varepsilon - \dot{M}$ relation.
As the trend is strongest for $\dot{M}$ one can speculate that
the quasar UV variability is related primarily to the accretion rate.
However, caution is required because the differences from the tests (correlation test, KS test)
do not strongly favour either of the three correlations. One also has to take into account
that $L$ is directly observed whereas $\dot{M}$ and $\varepsilon$ are derived quantities with explicit
dependence on $L$. An anti-correlation between
$V_m$ and $\dot{M}$ is qualitatively expected for the standard AD, but the expected slope of this relation
is much weaker than the observed one and, even more important, such an explanation is faced with the
time-scale problem (see above). 
\item[5.]
An anti-correlation between variability and accretion rate is qualitatively expected in the scenario
of strongly inhomogeneous ADs. This concept provides an interesting modification of the standard AD model
that has been shown to be able to explain some of the discrepancies between observations and
predictions for the standard AD (Dexter \& Agol \cite{Dexter11}), including the stochastic nature
of the variability. It would be interesting to see whether proper modelling can provide an adequate 
description of both the observed $V_m-\dot{M}$ relation and the strong variability in the far UV.
\end{itemize}

%%%%%%%%%%%%%%%%%%%%%%%%%%%%%%%%%%%%%%%%%%%%%%%%%%%%%%%%%%%%%%%%%%%%%%%%%%%%%%%%%%%%%%%%%%%%%%%%%%%%%%%%%%
%
\begin{acknowledgements}
We thank the anonymous referee for his useful comments that improved
the quality of the paper.
This research has made use of data products from the Sloan
Digital Sky Survey (SDSS). Funding for the SDSS and SDSS-II has been provided by
the Alfred P. Sloan Foundation, the Participating Institutions
(see below), the National Science Foundation, the National
Aeronautics and Space Administration, the U.S. Department
of Energy, the Japanese Monbukagakusho, the Max Planck
Society, and the Higher Education Funding Council for
England. The SDSS Web site is http://www.sdss.org/.
The SDSS is managed by the Astrophysical Research
Consortium (ARC) for the Participating Institutions.
The Participating Institutions are: the American
Museum of Natural History, Astrophysical Institute
Potsdam, University of Basel, University of Cambridge
(Cambridge University), Case Western Reserve University,
the University of Chicago, the Fermi National
Accelerator Laboratory (Fermilab), the Institute
for Advanced Study, the Japan Participation Group,
the Johns Hopkins University, the Joint Institute
for Nuclear Astrophysics, the Kavli Institute for
Particle Astrophysics and Cosmology, the Korean
Scientist Group, the Los Alamos National Laboratory,
the Max-Planck-Institute for Astronomy (MPIA),
the Max-Planck-Institute for Astrophysics (MPA),
the New Mexico State University, the Ohio State
University, the University of Pittsburgh, University
of Portsmouth, Princeton University, the United
States Naval Observatory, and the University of
Washington. 
\end{acknowledgements}
%
%%%%%%%%%%%%%%%%%%%%%%%%%%%%%%%%%%%%%%%%%%%%%%%%%%%%%%%%%%%%%%%%%%%%%%%%%%%%%%%%%%%%%%%%%%%%%%%%%%%%%%%%%%

%_________________________________________________________________________________________________________

%%%%%%%%%%%%%%%%%%%%%%%%%%%%%%%%%%%%%%%%%%%%%%%%%%%%%%%%%%%%%%%%%%%%%%%%%%%%%%%%%%%%%%%%%%%%%%%%%%%%%%%%%%
%
%
{}

\end{document}